\documentclass[nohyper,12pt,letterpaper]{JHEP}

\newfont{\frak}{eufm10 scaled 1200}

\newfont{\Bbb}{msbm10 scaled 1200}     
\newcommand{\mathbb}[1]{\mbox{\Bbb #1}}
\DeclareSymbolFont{AMSa}{U}{msa}{m}{n}
\DeclareSymbolFont{AMSb}{U}{msb}{m}{n}
\let\Box\relax
\DeclareMathSymbol{\Box}{\mathord}{AMSa}{"03}

\DeclareMathSymbol{\square}{\mathord}{AMSa}{"03}
\DeclareMathSymbol{\rightsquigarrow}{\mathrel}{AMSa}{"20}
\renewcommand{\Box}{\square}

\newdimen\tableauside\tableauside=1.0ex
\newdimen\tableaurule\tableaurule=0.4pt
\newdimen\tableaustep
\def\phantomhrule#1{\hbox{\vbox to0pt{\hrule height
\tableaurule width#1\vss}}}
\def\phantomvrule#1{\vbox{\hbox to0pt{\vrule width
\tableaurule height#1\hss}}}
\def\sqr{\vbox{%
  \phantomhrule\tableaustep
  \hbox{\phantomvrule\tableaustep\kern\tableaustep
\phantomvrule\tableaustep}%
  \hbox{\vbox{\phantomhrule\tableauside}\kern-\tableaurule}}}
\def\squares#1{\hbox{\count0=#1\noindent\loop\sqr
  \advance\count0 by-1 \ifnum\count0>0\repeat}}
\def\TT#1{\vcenter{\offinterlineskip
  \tableaustep=\tableauside\advance\tableaustep by-\tableaurule
  \kern\normallineskip\hbox
    {\kern\normallineskip\vbox
      {\gettableau#1 0 }%
     \kern\normallineskip\kern\tableaurule}%
  \kern\normallineskip\kern\tableaurule}}
\def\gettableau#1 {\ifnum#1=0\let\next=\null\else
  \squares{#1}\let\next=\gettableau\fi\next}

\def\IZ{{\mathbb Z}}

\def\6{\left({\bf 6}\right)}
\def\rseven{\left({\bf 7}\right)}
\def\rten{\left({\bf 10}\right)}

\def\27{\left({\bf 27}\right)}
\def\b27{\left(\overline{\bf 27}\right)}
\def\56{\left({\bf 56}\right)}
\def\78{\left({\bf 78}\right)}
\def\133{\left({\bf 133}\right)}

\newcommand{\drawsquare}[2]{\hbox{%
\rule{#2pt}{#1pt}\hskip-#2pt
\rule{#1pt}{#2pt}\hskip-#1pt
\rule[#1pt]{#1pt}{#2pt}}\rule[#1pt]{#2pt}{#2pt}\hskip-#2pt 
\rule{#2pt}{#1pt}}

\newcommand{\drawrule}[2]{\hbox{%
\rule[#1pt]{#1pt}{#2pt}}\rule[#1pt]{#2pt}{#2pt}\hskip-#2pt
}

\newcommand{\Yone}{\raisebox{6pt}{\drawsquare{6.5}{0.4}}}
\newcommand{\Ytwo}{\raisebox{-0.5pt}{\drawsquare{6.5}{0.4}}\hskip-6.9pt%
        \raisebox{6pt}{\drawsquare{6.5}{0.4}}}
\newcommand{\Ythree}{\raisebox{-7.0pt}{\drawsquare{6.5}{0.4}}\hskip-6.9pt%
        \raisebox{-0.5pt}{\drawsquare{6.5}{0.4}}\hskip-6.9pt
        \raisebox{6pt}{\drawsquare{6.5}{0.4}}}
\newcommand{\Yfour}{\raisebox{-13.5pt}{\drawsquare{6.5}{0.4}}\hskip-6.9pt%
        \raisebox{-7.0pt}{\drawsquare{6.5}{0.4}}\hskip-6.9pt
        \raisebox{-0.5pt}{\drawsquare{6.5}{0.4}}\hskip-6.9pt
        \raisebox{6pt}{\drawsquare{6.5}{0.4}}}

\newcommand{\Zone}{\raisebox{0pt}{\drawsquare{4.0}{0.4}}}
\newcommand{\Ztwo}{\raisebox{-4.0pt}{\drawsquare{4.0}{0.4}}\hskip-4.4pt%
        \raisebox{0pt}{\drawsquare{4.0}{0.4}}}

\newcommand{\Xone}{\raisebox{18.0pt}{\drawsquare{6.5}{0.4}}}%
\newcommand{\XBone}{\raisebox{20.0pt}{\drawrule{6.5}{0.4}}\hskip-6.9pt
	\raisebox{18.0pt}{\drawsquare{6.5}{0.4}}}%
\newcommand{\Xtwo}{\raisebox{11.5pt}{\drawsquare{6.5}{0.4}}\hskip-6.9pt
        \raisebox{18.0pt}{\drawsquare{6.5}{0.4}}} %
\newcommand{\Xthree}{
        \raisebox{5.0pt}{\drawsquare{6.5}{0.4}}\hskip-6.9pt
        \raisebox{11.5pt}{\drawsquare{6.5}{0.4}}\hskip-6.9pt
        \raisebox{18.0pt}{\drawsquare{6.5}{0.4}}} %
\newcommand{\Xfour}{
        \raisebox{-1.5pt}{\drawsquare{6.5}{0.4}}\hskip-6.9pt
        \raisebox{5.0pt}{\drawsquare{6.5}{0.4}}\hskip-6.9pt
        \raisebox{11.5pt}{\drawsquare{6.5}{0.4}}\hskip-6.9pt
        \raisebox{18.0pt}{\drawsquare{6.5}{0.4}}} %
\newcommand{\Xfive}{
        \raisebox{-8.0pt}{\drawsquare{6.5}{0.4}}\hskip-6.9pt
        \raisebox{-1.5pt}{\drawsquare{6.5}{0.4}}\hskip-6.9pt
        \raisebox{5.0pt}{\drawsquare{6.5}{0.4}}\hskip-6.9pt
        \raisebox{11.5pt}{\drawsquare{6.5}{0.4}}\hskip-6.9pt
        \raisebox{18.0pt}{\drawsquare{6.5}{0.4}}} %
\newcommand{\Xsix}{
        \raisebox{-14.5pt}{\drawsquare{6.5}{0.4}}\hskip-6.9pt
        \raisebox{-8.0pt}{\drawsquare{6.5}{0.4}}\hskip-6.9pt
        \raisebox{-1.5pt}{\drawsquare{6.5}{0.4}}\hskip-6.9pt
        \raisebox{5.0pt}{\drawsquare{6.5}{0.4}}\hskip-6.9pt
        \raisebox{11.5pt}{\drawsquare{6.5}{0.4}}\hskip-6.9pt
        \raisebox{18.0pt}{\drawsquare{6.5}{0.4}}} %
\newcommand{\Xseven}{
        \raisebox{-21.0pt}{\drawsquare{6.5}{0.4}}\hskip-6.9pt
        \raisebox{-14.5pt}{\drawsquare{6.5}{0.4}}\hskip-6.9pt
        \raisebox{-8.0pt}{\drawsquare{6.5}{0.4}}\hskip-6.9pt
        \raisebox{-1.5pt}{\drawsquare{6.5}{0.4}}\hskip-6.9pt
        \raisebox{5.0pt}{\drawsquare{6.5}{0.4}}\hskip-6.9pt
        \raisebox{11.5pt}{\drawsquare{6.5}{0.4}}\hskip-6.9pt
        \raisebox{18.0pt}{\drawsquare{6.5}{0.4}}} %

\def\Xn#1{ \scriptstyle{#1}\left\lbrace{
        \raisebox{-21.0pt}{\drawsquare{6.5}{0.4}}\hskip-6.9pt
        \raisebox{-8.0pt}{\vdots}\hskip-3.2pt
        \raisebox{5.0pt}{\drawsquare{6.5}{0.4}}\hskip-6.9pt
        \raisebox{11.5pt}{\drawsquare{6.5}{0.4}}\hskip-6.9pt
        \raisebox{18.0pt}{\drawsquare{6.5}{0.4}} } \right. } %
\def\XBn#1{ \scriptscriptstyle{#1}\left\lbrace{ 
	\raisebox{20.0pt}{\drawrule{6.5}{0.4}}\hskip-6.9pt
        \raisebox{-21.0pt}{\drawsquare{6.5}{0.4}}\hskip-6.9pt
        \raisebox{-8.0pt}{\vdots}\hskip-3.2pt
        \raisebox{5.0pt}{\drawsquare{6.5}{0.4}}\hskip-6.9pt
        \raisebox{11.5pt}{\drawsquare{6.5}{0.4}}\hskip-6.9pt
        \raisebox{18.0pt}{\drawsquare{6.5}{0.4}} } \right. } %

\def\XN#1{ \left. \raisebox{-27.5pt}{\drawsquare{6.5}{0.4}}\hskip-6.9pt
        \raisebox{-21.0pt}{\drawsquare{6.5}{0.4}}\hskip-6.9pt
        \raisebox{-8.0pt}{\vdots}\hskip-3.2pt
        \raisebox{5.0pt}{\drawsquare{6.5}{0.4}}\hskip-6.9pt
        \raisebox{11.5pt}{\drawsquare{6.5}{0.4}}\hskip-6.9pt
        \raisebox{18.0pt}{\drawsquare{6.5}{0.4}}\right\rbrace  
\scriptstyle{#1} } %

\def \b{\beta}

\title{Spectroscopy of Gauge Theories Based on Exceptional Lie Groups}
\author{Philippe Pouliot\\
 Physics Department\\
University of Texas at Austin\\
Austin, TX 78712 USA\\
Email: \email{pouliot@physics.utexas.edu}}
\abstract{We generate by computer a basis of invariants
for the fundamental representations of the exceptional 
Lie groups $E_6$ and $E_7$, up to degree 18.
We discuss the relevance of this calculation for the study of 
supersymmetric gauge theories, and revisit the self-dual exceptional models.
We study the chiral ring of $G_2$ to degree 13, as well as a few classical 
groups. 
The homological dimension of a ring is a natural estimator of its 
complexity and 
provides a guideline for identifying theories that have a good chance 
to be amenable to 
a solution.}
\keywords{Supersymmetry, Chiral Rings, Gauge Theories, Duality, 
LiE, Syzygy, Classical Invariant Theory}
\preprint{UTTG-10-01}

\begin{document}

\newpage
\section{Introduction}
In the past several years, much progress has been made in understanding 
the behavior of supersymmetric theories at low energy.
In the best of cases, weakly coupled dual descriptions have been found 
for strongly coupled supersymmetric theories, 
which is tantamount
to an exact solution of the theory in the very low energy regime and 
for very large distances\footnote{For practical reasons,
we will arbitrarily limit
our context to ${\cal N}=1$ supersymmetric gauge theories in four 
dimensions, although our results have a more general range of 
applicability and
although 
there has been considerable work relevant to the issues addressed 
in this paper coming from string theory, extended supersymmetry, or
theories in more or fewer dimensions.}. 

However, many key questions have not been answered. For example, 
given a gauge group and matter content superfields in some
representation of the gauge group, what is the low energy behavior 
of the theory? For starters, most theories are free in the low
energy regime since they are not asymptotically free. Of those that 
are asymptotically free, 
most live in an interacting non-Abelian Coulomb phase.
A smaller fraction have extended supersymmetry or live in a confining phase 
or an Abelian Coulomb phase at low energies, and are rather well understood.
But the behavior of the majority of theories in a non-Abelian Coulomb 
phase remains to be understood. Although not of much direct physical 
interest,
it is an important mathematical physics problem. 

An important insight into the behavior of theories in a non-Abelian 
Coulomb phase is duality~\cite{Seiberg:1995pq}. It is unclear
whether duality is generic or a feature of a few especially simple 
theories. Thus a great deal of work was done and 
has led to finding 
many more examples of duality
beyond the examples of~\cite{Seiberg:1995pq} 
for the classical groups $SU$, $SO$ and $Sp$ with matter fields
in copies of the fundamental representations. The simplest examples of 
duality arise when the theory confines and 
the low energy dual description just consists of gauge invariant mesons 
and baryons
\cite{Seiberg:1994bz}--\cite{Klein:1999mj} and there is a claim that the 
theories in a confining phase 
have all been found and studied, at least for simple gauge 
groups~\cite{Dotti:1998rv,Dotti:1998wn}. 
When the dual theory is not confining, there are
numerous examples for the classical groups with a variety of features: 
terms in the electric superpotential, 
tensor products of groups, (anti)-symmetric or adjoint representations, 
finite theories,~\cite{Kutasov:1995ve}--\cite{Karch:1998ck},
while for the exceptional groups, there is a family
of examples based on $G_2$ and the $Spin$ 
groups~\cite{Pouliot:1995zc}--\cite{Kawano:1996bd}
and isolated examples of a 
so-called ``self-duality''~\cite{PouliotStrassler}--\cite{Cho:1998am}.

The battery of tests that known dualities pass are also 
the tools being used to search for new examples. 
Several of these tests require detailed knowledge of the 
theory being studied. This is the case for matching the flat 
directions, matching the spectra or the chiral rings, 
or for checking 't Hooft's anomaly matching 
conditions~\cite{'tHooft:1980xb}. For all these tests, 
a knowledge of the gauge invariant chiral superfields is necessary.

Among the many impediments in finding more examples, one problem 
that is quite tractable is obtaining the structure of the chiral 
rings for theories of interest. 
In this paper, we extend the amount of data known about such gauge 
invariant superfields for some specific cases.  
We focus our effort on the exceptional simple Lie groups $E_6$ 
and $E_7$ with matter quarks in many copies of the fundamental
representation.
We also illustrate with other groups how very detailed information 
can be obtained, systematically and for any representation of any Lie group, 
with enough computer power.
In the process, we
make tools, which are not new, more easily available, for generating 
this kind of data.

Now a short summary of the contents. In section~2, we provide a brief 
mathematical guide and references.
In section~3, we mention the existence of 2 theories which must have a 
dual description, which offers some 
motivation for the calculations done in this paper.
In section~4, we find an (almost) complete list of the invariants of 
$E_6$ for copies of the $27$-dimensional
representation, a list of $20$ invariants up to degree 18. 
In section~5, we show how this effort stalls for $E_7$, by exhibiting 
several hundred invariants, with many more
yet to be found. 
The present author sees this complexity as a reasonable apology for 
failing to find the dual descriptions.
We then move on to simple examples and recover some known mathematical 
results.
In section~6.1, we study in some detail the syzygy chain for $SU(2)$ 
with fundamentals; in section~6.2,
we recover the well-known results for $G_2$; in sections 6.3, 6.4 
and 6.5, we find some invariants for copies the 
symmetric tensor representations of $SU(3)$, $SU(4)$ and $SU(5)$, 
and observe that the complexity dramatically
increases with the rank of the group. 
We include appendices on constraints and glueballs in $E_6$, as well 
as some computer code for the LiE software that we
used to do these calculations.
In the conclusion, we mention some obvious extensions to this work, 
and briefly discuss its applicability.

\section{Mathematical Preliminaries}

Our goal in this paper is to find a minimal list of 
the ``fundamental'' polynomial 
invariants\footnote{For a section of the mathematical 
literature relevant to invariant theory:~\cite{Weyl:1946}--\cite{Gufan:2001}.
}. This is known as a Hilbert basis:
a finite collection of invariants $I_1,\ldots,I_m$ forms a 
Hilbert basis if every other invariant can be written as a 
polynomial function of the basis
invariants: $I=P(I_1,\ldots,I_m)$ (\cite{Olver:1999}, page 39.). 
An important theorem of Hilbert showed that 
any finite system of homogeneous polynomials admits a Hilbert basis. 
The elements of the Hilbert basis are said to be (polynomially) 
independent.

There are other, less stringent, notions of independence: the 
invariants can be rationally, algebraically, or functionally independent. 
For the purposes of duality for supersymmetric theories in a 
non-Abelian Coulomb phase,
which notion of independence is relevant? At the most basic level 
of matching flat directions between the electric and the magnetic theories, 
all that matters is functional independence: that the moduli spaces 
of vacuum states have the same dimensions and that the theories remain dual
along the flat directions. 
At the level of 't Hooft anomaly matching, functional independence is 
clearly not enough 
and perhaps algebraic independence is what one is asking for: 
if 't Hooft's anomaly matchings are satisfied for a basic set of 
invariants, they would no longer be 
satisfied for the invariants obtained by acting
with some functions on these invariants. 
Finally, the best one can impose from the requirement of supersymmetry 
is a complete isomorphism of the chiral 
rings~\cite{Kutasov:1996ss,Brax:1999gy,Pouliot:1999yv}. For other theoretical
work relevant to these issues, 
see \cite{Buccella:1982nx}--\cite{Brax:2001an}.

The chiral rings that arise in supersymmetric gauge 
theories can be very simple or extremely complicated. 
In ring theory, there is a natural notion of the complexity of a 
ring, measured by its homological dimension.
The invariants typically satisfy constraints, called first-order 
syzygies. These first-order syzygies themselves satisfy constraints, called
second-order syzygies, and so forth. For the rings that we are 
concerned with here, Hilbert's theorem applies, and this chain 
of syzygies must terminate. 
The length of this chain is known as the homological dimension. 

When the ring is freely generated, the homological dimension is 
zero. However, when it is not freely generated, it is typically very large.
That these rings are very complicated is known by mathematicians. 
For example, 
for {\sl irreducible} representations other than the fundamental or 
the adjoint, the homological dimension of $E_7$ is known to be larger 
than $26334$ 
(page 11 of \cite{popov:1992})!, with similar surprisingly large lower 
bounds on the homological dimensions for other exceptional and spin groups.
However, in this paper, we are mostly concerned with multiple copies 
of the fundamental representation. We will find that in these cases too, 
the homological
dimension is likely to be very large.

\section{Theories that must have a dual}
Although this complexity was known to some physicists 
(cf. page 2 of \cite{Kutasov:1996ss}), it had escaped the author.
Most supersymmetric gauge theories exist in a non-Abelian Coulomb 
phase at long distances. Among them,
just a few have the noteworthy feature that a gauge invariant baryon 
in their spectrum has an $R$-charge that is less than $2/3$.
Such theories must have a dual description~\cite{Seiberg:1995pq}, 
because the spectrum of the electric theory is not in a unitary 
representation of the superconformal algebra.
Furthermore, this baryon of charge less than $2/3$ must appear as a 
fundamental 
free field in the dual description. It is also to be expected, if the 
known examples can be a guide,
that the full dual description will be free.
Below, we give examples of such theories that must have duals. 
Our analysis of these examples is inconclusive, but might possibly 
be of interest to someone seeking to find duals.  

\subsection{$E_6$ with 5 flavors}
One such example~\cite{Pouliot:1996ab,Cho:1998am} 
is the theory with ${\cal N}=1$ supersymmetry and with gauge group 
$E_6$ and five flavors in the $27$-dimensional fundamental representation.
The global anomaly-free symmetry is $SU(5)\times U(1)_R$. In the table 
below, we list the quantum numbers of the fields.
\begin{center}
\begin{tabular}{|c|c|c|c|}
\hline
& $E_{6}$ & $SU(5)$ & $U(1)_R$ \\
\hline
$Q$ & $27$ & $\Zone$ & $1/5$ \\
\hline\hline
$I_1$ & 1 & $\Zone\Zone\Zone$  & $3/5$\\
$I_2$ & 1 & $\overline{\Ztwo\Ztwo}$ & $6/5$ \\
$I_3$ & 1 & $\Ztwo\Ztwo$  & $9/5$ \\
$I_4$ & 1 & $\overline{\Zone\Zone\Zone}$ & $12/5$ \\
\hline
\end{tabular}
\end{center}
We have included the spectrum of polynomially independent gauge 
invariant baryons. 
This theory is in a non-Abelian Coulomb phase, since $SU(3)$ along 
its flat direction is. 
And because the $R$ charge of $I_1$ is less than $2/3$, this theory 
must have a dual.

For future reference, we give the
contributions of the various fields to the anomalies:
\begin{center}
\begin{tabular}{|c|c|c|c||c|c|c|c|}
\hline
Gauge & Field & $SU(5)$ & $U(1)_R$ & $SU(5)^3$ & $SU(5)^2U(1)_R$ & 
$U(1)_R$ & $ U(1)_R^3$ \\
\hline
$27$ & $Q$ & $\Zone$ & $1/5$ & $27$ & $-108/5$ & $-108$ & $-1728/25$ \\
$78$ & $\lambda$ & 1 & 1 & 0 & 0 & $78$ & $78=1950/25$ \\
 & Total & & & $27$ & $-108/5$ & $-30$ & $222/25$ \\
\hline
\hline
&$I_1$ & $\Zone\Zone\Zone$ & $3/5$ & $44$ & $-56/5$ & $-14$ & $-56/25$ \\
&$I_2$ & $\overline{\Ztwo\Ztwo}$ & $6/5$ & $-15$ & $42/5$ & $10$ & $2/5$ \\
&$I_3$ & $\Ztwo\Ztwo$ & $9/5$ & $15$ & $168/5$ & $40$ & $128/5$ \\
&$I_4$ & $\overline{\Zone\Zone\Zone}$ & $12/5$ & $-44$ & $196/5$ & $49$ 
& $2401/25$ \\
&$W^2_\alpha Q^3$ & $\Zone\Zone\Zone$ & $13/5$ & $44$ & $224/5$  & 56 
& $3584/25$\\
&$W^2_\alpha Q^3$ & $\Ztwo\Zone$ & $13/5$ & $16$ & $176/5$ & $64$ 
& $4096/25$ \\
&$W^2_\alpha Q^3$ & $\overline{\Ztwo}$ & $13/5$ & $-1$ & $16$ & $64$ 
& $1024/25$ \\
&$W^2_\alpha Q^6$ & $\Zone$ & $16/5$ & $1$ & $11/5$ & $11$ & $1331/25$ \\
\hline
\end{tabular}
\end{center}
We will now make a few observations, encountered during the search for 
a dual. To begin, 
we will make the hypothesis that there does exist a free dual description. 
This implies that the $SU(5)$ global symmetry is explicitly realized 
and that $I_1$ is the 
only gauge singlet which is an elementary field in the dual. 
Thus, the dual description found for $Spin(8)$ in
\cite{Berkooz:1997bb, Cho:1998am} will not be helpful here. 
In particular, this dual description contains several gauge-singlets 
transforming as $5$-dimensional $4$-index symmetric tensors
of their global $SU(2)$. 
These singlets cannot all come from the free baryon $I_1$, but must 
come from $I_2$ or $I_3$. This means that 
their example, although an impressive accomplishment by itself, 
has not been ``fully'' dualized.

One key problem in narrowing the search for a dual for $E_6$ with 5 
flavors is understanding how to generate the baryons $I_2$, $I_3$, 
$I_4$ without generating undesirable
invariants at the same time. 
Another clue is the matching of the 't Hooft anomaly for $SU(5)^3$. 
The $27$ coming from the electric theory is badly matched by 
the $44$ coming from the baryon $I_1$. Where do the $-17$ come from? 
$-17$ here is a rather large number in this context. 
One could imagine a decomposition $17=2+3\cdot 5=5+3\cdot 4$. 
This would mean three gauge groups in the dual, with fields transforming 
under 
$G_1\times G_2 \times G_3\times SU(5)$ as $(2,1,1,
\overline{5})\oplus (1,3,5,\overline{5})$ or as 
$(5,1,1,\overline{5})\oplus (1,3,4,\overline{5})$. Such examples 
generate lots of unwanted invariants. 
To make progress, one would have to develop an understanding of how 
a superpotential can remove such unwanted invariants.

One simple way to avoid generating lots of unwanted invariants is to 
have $I_2$, $I_3$ as fundamental fields in the dual, 
but transforming under a $U(1)$ gauge symmetry.  
This is not necessarily an unattractive possibility, since free dual 
descriptions are typically not asymptotically free. 
However, this provides little help in cancelling
the $SU(5)^3$ anomaly. One can wonder if the techniques 
of~\cite{Argyres:1996eh} could be of more general applicability and
allow us to get information on what the free dual quarks could 
be, or at least what the rank or dimension of the dual
gauge group is. Perhaps string theory constructions could also 
shed light on this problem.

Another curious feature is that the invariants of $SU(3)$ with $5$ 
symmetric tensors are very similar to
the invariants of $E_6$ with $5$ fundamentals: namely, $I_1$, $I_2$, 
and $I_3$ are identical, while
$I_4$ differs. The constraints among these invariants, and the chiral 
rings, and the glueballs,
are of course different.
It is not clear whether this observation has any significance.

\subsection{$Spin(16)$ with one spinor and two vectors}
This is another example of an asymptotically free theory which is in 
a non-Abelian Coulomb phase at long distances, and which must have a 
dual description.
$Spin(16)$ has two real $128$-dimensional spinor representations of 
opposite chiralities. Let's consider the theory with just one spinor 
$Q=128_+$,
and with $N$ vectors $V=16$. There is a choice of the $R$-charge for 
which the $R$-charge of $Q$ and $V$ are the same, and equal to
$1-\frac{7}{8+N}$. This is appropriate, since both $Q^2$ and $V^2$ are 
invariants.  

For $N=0$, the theory confines~\cite{Dotti:1998wn}, with the  
spectrum of invariants $Q^2$, $Q^8$, $Q^{12}$, $Q^{14}$, $Q^{18}$, 
$Q^{20}$, $Q^{24}$, $Q^{30}$, as follows from 
knowledge of the second Casimir invariants of $E_8$. We do not 
consider the case $N=1$ here.
For $N=2$, along the $Spin(14)$ flat direction, $Q$ breaks into 
$64+64'$. One of the $64$ then Higgses
$Spin(14)$ to $G_2\times G_2$~\cite{Dotti:1998wn}. The remaining 
$64$-dimensional spinor
presumably decomposes into $(7,7)\oplus (7,1)\oplus (1,7)\oplus (1,1)$,
and these $G_2$ theories are known to be in a non-Abelian Coulomb phase. 
Since
$Q^2$ and $V^2$ have $R$-charge equal to $3/5$, which is less than 
$2/3$, the $Spin(16)$ theory with $N=2$ must have a dual description.
The difficulties in finding a dual for this theory 
are clearly of a very different nature than for $E_6$ and we will have 
nothing further
to say about it here. 

\section{Invariants of $E_6$ with fundamentals}

Our main result in this paper is a list of the polynomially independent 
invariants of 
multiple copies of the fundamental representations of
 $E_6$ and $E_7$, up to degree 18. 
We find it convenient to express our results in the context of 
supersymmetric gauge theories,
even though they have a more general range of application.
For $E_6$ with chiral superfields in the $27$-dimensional fundamental 
representation, 
three invariants were previously well-known, at least in the gauge 
theory community:
\begin{eqnarray}
I_1=\27^3_{[3]}=\  \Yone\Yone\Yone \qquad 
I_2=\27^6_{[2^3]}=\  \Ythree\Ythree \qquad 
I_4=\27^{12}_{[3^4]}=\  \Yfour\Yfour\Yfour\nonumber\ .
\end{eqnarray}
However, an invariant of lower degree than $I_4$, of degree 9, had 
attracted little attention:
\begin{eqnarray}
I_3=\27^{9}_{[1^5\,2^2]}=\ \Xfive\Xtwo\Xtwo\nonumber\ .
\end{eqnarray}
Previously, it had been unclear whether
\begin{eqnarray}
\Xsix\nonumber
\end{eqnarray}
should be included as an independent invariant of $E_6$. We find 
here that $E_6$ does not have such an invariant. 
It is however an independent invariant of the symmetric tensor of 
$SU(3)$, as stated below in the
section on $SU(3)$.
We found two more invariants of degree 12:
\begin{eqnarray} 
I_5&=&\27^{12}_{[1^6\,3^2]}=\ \Xsix\Xtwo\Xtwo\Xtwo \qquad
I_6=\27^{12}_{[1^7\,1^3\,2]}=\ \Xseven\Xthree\Xone\Xone \nonumber\ .
\end{eqnarray}
And in degree 15, we found five more invariants:
\begin{eqnarray}
I_7&=&\27^{15}_{[1^8\, 1^3\, 1^2\, 2]}=\ \Xn8\Xthree\Xtwo\Xone\Xone \qquad 
I_{8}=\27^{15}_{[1^9\, 1^4\, 2]}=\ \Xn9\Xfour\Xone\Xone \qquad 
I_9=\27^{15}_{[1^9\, 2^3]}=\ \Xn9\Xthree\Xthree \nonumber\  \\
I_{10}&=& \27^{15}_{[1^9\, 2^2\, 2]}=\ \Xn9\Xtwo\Xtwo\Xone\Xone \qquad
I_{11}=\27^{15}_{[1^{10}\, 5]}=\ \Xn{10}\Xone\Xone\Xone\Xone\Xone\nonumber\ .
\end{eqnarray}
There were no constraints among the invariants up to degree 12, but 
five constraints (first-order syzygies) arise in degree 15:
\begin{eqnarray}
C_{1}^{(15)} &=& \left( I_1I_4+I_2I_3 =0\right)_{[2^5\, 1^4\, 1]} =\  
\Xfive\Xfive\Xfour\Xone\nonumber\ \\
C_{2}^{(15)} &=& \left( I_1I_2^2+I_1^2I_3+I_1I_4+I_2I_3=0\right)_{[1^5\, 
2^4\, 2]} =\ \Xfive\Xfour\Xfour\Xone\Xone \nonumber\  \\
C_{3}^{(15)} &=& \left( I_2I_3 =0\right)_{[1^6\, 1^5\, 1^4]}=\  
\Xsix\Xfive\Xfour \nonumber\ \\
C_{4}^{(15)} &=& \left( I_1^2I_3+I_1I_2^2+I_1I_5+I_2I_3=0 
\right)_{[1^6\, 2^3\, 1^2\, 1]} =\  \Xsix\Xthree\Xthree\Xtwo\Xone 
\nonumber\ \\
C_{5}^{(15)} &=& \left( I_2I_3+I_1I_6 =0\right)_{[1^7\, 1^4\, 2^2]} =\  
\Xseven\Xfour\Xtwo\Xtwo \nonumber\ .
\end{eqnarray} 
In these expressions for the constraints, we indicate that a linear 
combination of invariants is constrained, along with the Young tableau 
describing how the
indices are meant to be contracted. We did not check whether some of 
the coefficients in these linear combinations could be zero.  
And in degree 18, there are nine more invariants:
\begin{eqnarray}
I_{12} &=& \27^{18}_{[1^{10}\,1^4\,1^2\,2]}=\ 
\Xn{10}\Xfour\Xtwo\Xone\Xone\qquad
I_{13} = \27^{18}_{[1^{10}\,2^3\,1^2]}=\ \Xn{10}
\Xthree\Xthree\Xtwo\nonumber\ \\
I_{14} &=& \27^{18}_{[1^{10}\,1^3\,1^2\,3]}=\ 
\Xn{10}\Xthree\Xtwo\Xone\Xone\Xone\qquad
I_{15} = \27^{18}_{[1^{11}\,1^5\,1^2]}=\ \Xn{11}\Xfive\Xtwo\nonumber\ \\
I_{16} &=& \27^{18}_{[1^{11}\,1^4\,1^2\,1]}=\ \Xn{11}\Xfour\Xtwo\Xone\qquad
I_{17} = \27^{18}_{[1^{11}\,1^3\,2^2]}=\ \Xn{11}
\Xthree\Xtwo\Xtwo\nonumber\ \\
I_{18} &=& \27^{18}_{[1^{11}\,1^3\,1^2\,2]}=\ \Xn{11}
\Xthree\Xtwo\Xone\Xone\qquad 
I_{19} = \27^{18}_{[1^{11}\,1^3\,4]}=\ \Xn{11}
\Xthree\Xone\Xone\Xone\Xone\nonumber\ \\
I_{20} &=& \27^{18}_{[1^{12}\,1^3\,3]}=\ \Xn{12}
\Xthree\Xone\Xone\Xone\nonumber\ .
\end{eqnarray}
There are many constraints in degree 18, and we list 
them in an appendix. It is possible that this
list of 20 invariants is not an exhaustive list of invariants. 
It would take over 50 days of computer time on a Pentium 4 
machine to get the invariants of degree 21, using 1 GB of RAM,
without improvement in the group theory software.
We expect however that our list is almost complete. 
In fact, we seem to have all the invariants required for the 
subgroups of $E_6$. Along the $SO(10)$ flat direction for 
example, we ought to find
a completely antisymmetrized baryon for the 10-dimensional 
vector: $({\bf 10}^{10})=[1^{10}]$. This invariant of $SO(10)$ 
must come from the invariant $I_{11}$
of $E_6$.
As usual, information can be inferred about the invariants of 
the theories along the flat directions of $E_6$ from the
knowledge of the basic invariants $I_1$ through $I_{20}$.

An understanding of the glueball invariants is also essential 
for the study of duality. This is particularly relevant for 
$E_6$ since it has $SO(N)$ subgroups,
for which the glueballs play a crucial role in the matching of 
invariants~\cite{Intriligator:1995id}. Unfortunately, 
there are so many glueball invariants that we chose not to study
the situation in more detail. Some results are to be found in the appendix. 
 
\subsection{Status of the self-dual model for $E_6$ with 6 flavors}

In \cite{Cho:1998am}, Cho studied in detail the flat directions 
of the $E_6$ self-dual model of \cite{PouliotStrassler,
Ramond:1997ku,Distler:1997ub}.
He considered the model along the electric $Spin(8)$ flat 
direction, and found an invariant $d''$ of the $E_6$ dual theory with the 
quantum numbers $(\TT{1},5/2)$ under the global 
symmetry $SU(4)\times U(1)_R$. However, we point out that 
there are 4 invariants in the tensor product $8_v^3\otimes 8_s\otimes 8_c$,
4 invariants in $8_s^3\otimes 8_v\otimes 8_c$ 
and 4 invariants in $8_c^3\otimes 8_v\otimes 8_s$ 
all with $R$-charge $5/2$. There is not a scarcity of invariants that
$d''$ could correspond to, although a more detailed 
analysis would be required to 
tell which of these 12 (not all independent) 
invariants is the right one. 
Thus we do not see this problem 
as invalidating this duality.

A potentially more serious problem is the mapping of the previously unknown 
invariant $I_3$. Corresponding to $I_3$, there is an invariant
\begin{eqnarray}
{\tilde {\cal I}}=\ \overline{\TT{3 2 2 2}}\ \nonumber\ .
\end{eqnarray}
However, $\tilde{\cal I}$ is not a basic invariant: 
it is simply the product of $\tilde I_1$ and $\tilde I_2$ from the
dual theory. Furthermore, the dual superpotential renders 
$\tilde I_1$ redundant. Could
$\tilde {\cal I}$ be considered to be a bona fide basic invariant? 

Another puzzle is that the dual theory has an invariant 
$\tilde I_3$. This corresponds to 
\begin{eqnarray}
{\cal I}=\ \TT{3 2 2 2}\ \nonumber\ 
\end{eqnarray} 
in the electric theory. ${\cal I}$ is certainly not a basic 
invariant, being the product of $I_1$ and $I_2$.
Therefore, we need to get rid of $\tilde I_3$. Perhaps 
$\tilde I_3$ is made redundant by the dual superpotential. 
That might force us to relax the requirements of duality 
to rational or algebraic independence of the invariants 
instead of the stronger
polynomial independence. 
Our attitude is that this self-dual model is complicated 
and that we do not have the technology to decisively confirm
or disprove this duality\footnote{There is another serious
problem with this self-dual model~\cite{Csaki:1998aw}, with the
$\IZ_{36}$ global symmetry. One way around this problem is
to add the invariant $I_2$ to the superpotential of both electric
and magnetic theories. The resulting global symmetry is
$Spin(6)\times U(1)_R\times \IZ_6$. We thank Andreas Karch for
reminding us.}.

\section{Invariants of $E_7$ with fundamentals}
We repeat the search for invariants of the previous section, 
but this time with the $56$-dimensional fundamental representation of $E_7$.
The well-known invariants are:
\begin{eqnarray}
I_1&=& \56^2_{[1^2]}=\ \Ytwo   \qquad I_2=\56^4_{[4]}=\ 
\Yone\Yone\Yone\Yone\nonumber\\
I_3&=&\56^6_{[3^2]} =\  \Ytwo\Ytwo\Ytwo \qquad
   I_4=\56^8_{[2^3\, 2]}=\ \Ythree\Ythree\Yone\Yone \nonumber\ .
\end{eqnarray}
We then find three invariants of degree 10:
\begin{eqnarray}
I_5&=&\56^{10}_{[2^4\,1^2]}=\ \Yfour\Yfour\Ytwo \qquad 
I_6=\56^{10}_{[1^4\,1^3\,3]}=\ \Yfour\Ythree\Yone\Yone\Yone\qquad
I_7=\56^{10}_{[1^4\,2^3]}=\ \Yfour\Ytwo\Ytwo\Ytwo  \nonumber\ .
\end{eqnarray}
Nine invariants of degree 12:
\begin{eqnarray}
I_8&=&\56^{12}_{[4^3]}=\ \Ythree\Ythree\Ythree\Ythree \qquad 
I_{9}=\56^{12}_{[2^4\,4]}=\ \Xfour\Xfour\Xone\Xone\Xone\Xone \qquad
I_{10}=\56^{9}_{[1^4\,1^3\,2^2\,1]}=\ 
\Xfour\Xthree\Xtwo\Xtwo\Xone\nonumber\ \\
I_{11}&=&\56^{12}_{[2^5\,2]}=\ \Xfive\Xfive\Xone\Xone\qquad
I_{12}=\56^{12}_{[1^5\,1^4\,1^2\,1]}=\ \Xfive\Xfour\Xtwo\Xone  \qquad
I_{13}=\56^{12}_{[1^5\,2^3\, 1]}=\ \Xfive\Xthree\Xthree\Xone  \nonumber\  \\
I_{14}&=&\56^{12}_{[1^5\,1^3\,1^2\,2]}=\ \Xfive\Xthree\Xtwo\Xone\Xone \qquad
I_{15}=\56^{12}_{[1^5\,2^2\,3]}=\ \Xfive\Xtwo\Xtwo\Xone\Xone\Xone \quad
I_{16}=\56^{12}_{[2^6]}=\ \Xsix\Xsix \nonumber\ .
\end{eqnarray}
We then find thirty invariants of degree 14, where we begin to find
a multiplicity of invariants with identical transformation properties.
For example, the computer counted a total of $14$ invariants 
with the Young tableau shape:
\begin{eqnarray}
\Xsix\Xthree\Xthree\Xone\Xone\nonumber\ ,
\end{eqnarray}
but then found that this Young tableau can be obtained in 
twelve different ways from 
the lower degree invariants\footnote{Namely
$I_1^5I_2$, $I_1^4I_3$, $I_1I_3^2$, $I_1 I_{13}$, 
$I_1I_{14}$, $2\times I_1^2I_2I_3$, $2\times I_1^2I_6$,
$3\times I_1^3 I_4$.}, leaving us with a count of $2$ new basic 
invariants with that
particular shape. Thus the new invariants of degree 14 are:
\begin{eqnarray}
\56^{14}_{[3^4\,1^2]}&=&\  \Yfour\Yfour\Yfour\Ytwo \qquad
\56^{14}_{[2^4\,1^3\,1^2\,1]}=\  \Yfour\Yfour\Ythree\Ytwo\Yone \nonumber\\
\56^{14}_{[2^4\,3^2]}&=&\  \Yfour\Yfour\Ytwo\Ytwo\Ytwo \qquad
\56^{14}_{[1^4\,2^3\,1^2\,2]}=\  \Yfour\Ythree\Ythree\Ytwo\Yone\Yone 
\nonumber
\end{eqnarray}
\begin{eqnarray} 
\56^{14}_{[2^5\,1^2\,2]}&=&\  \Xfive\Xfive\Xtwo\Xone\Xone  \qquad
\56^{14}_{[1^5\,2^4\,1]}=\  \Xfive\Xtwo\Xtwo\Xtwo\Xtwo\Xone \qquad
\56^{14}_{[1^5\,1^4\,1^3\,1^2]}=\  \Xfive\Xfour\Xthree\Xtwo \nonumber\\
\56^{14}_{[1^5\,1^4\,1^3\,2]}&=&\  \Xfive\Xfour\Xthree\Xone\Xone \qquad
\56^{14}_{[1^5\,1^4\,2^2\,1]}=\  \Xfive\Xfour\Xtwo\Xtwo\Xone  \qquad
\56^{14}_{[1^5\,1^4\,1^2\,3]}=\  \Xfive\Xfour\Xtwo\Xone\Xone\Xone  
\nonumber\ \\
\56^{14}_{[1^5\,2^3\,1^2\,1]}&=&\  \Xfive\Xthree\Xthree\Xtwo\Xone \qquad
\56^{14}_{[1^5\,2^3\,3]}=\  \Xfive\Xthree\Xthree\Xone\Xone\Xone 
\nonumber\ \\
\56^{14}_{[1^5\,1^3\,2^2\,2]}&=&\  \Xfive\Xthree\Xtwo\Xtwo\Xone\Xone  \qquad
\56^{14}_{[1^5\,1^3\,1^2\,4]}=\  
\Xfive\Xthree\Xtwo\Xone\Xone\Xone\Xone  \nonumber
\end{eqnarray}
\begin{eqnarray}
\56^{14}_{[2^6\,1^2]}&=&\ \Xsix\Xsix\Xtwo  \qquad
\56^{14}_{[1^6\,1^5\,1^2\,1]}=\  \Xsix\Xfive\Xtwo\Xone  \qquad
\56^{14}_{[1^6\,1^5\,3]}=\  \Xsix\Xfive\Xone\Xone\Xone \nonumber\ \\
\56^{14}_{[1^6\,2^4]}&=&\ \Xsix\Xfour\Xfour \qquad
\56^{14}_{[1^6\,1^4\,1^3\,1]}=\  \Xsix\Xfour\Xthree\Xone \qquad
2\times\ \56^{14}_{[1^6\,1^4\,2^2]}=\  \Xsix\Xfour\Xtwo\Xtwo \nonumber\ \\
\56^{14}_{[1^6\,1^4\,1^2\,2]}&=&\  \Xsix\Xfour\Xtwo\Xone\Xone \qquad
\56^{14}_{[1^6\,1^4\,4]}=\  \Xsix\Xfour\Xone\Xone\Xone\Xone \qquad
2\times\ \56^{14}_{[1^6\,2^3\,2]}=\  \Xsix\Xthree\Xthree\Xone\Xone 
\nonumber\ \\
\56^{14}_{[1^6\,1^3\,2^2\,1]}&=&\  \Xsix\Xthree\Xtwo\Xtwo\Xone \qquad
\56^{14}_{[1^6\,1^3\,1^2\,3]}=\  \Xsix\Xthree\Xtwo\Xone\Xone\Xone 
\nonumber\ \\
\56^{14}_{[1^6\,4^2]}&=&\  \Xsix\Xtwo\Xtwo\Xtwo\Xtwo \qquad
\56^{14}_{[1^6\,2^2\,4]}=\  \Xsix\Xtwo\Xtwo\Xone\Xone\Xone\Xone\nonumber
\end{eqnarray}
\begin{eqnarray}
\56^{14}_{[1^7\,1^5\,2]}&=&\  \Xseven\Xfive\Xone\Xone  \nonumber\ .
\end{eqnarray}
Then we find $125$ basic invariants of degree 16 (from here on, we find the 
Young tableau generating macro
written by J. Distler for~\cite{Distler:1997ub} extremely useful):
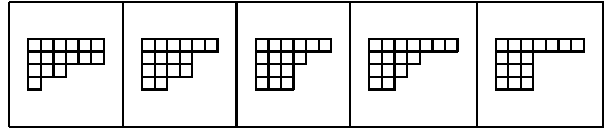
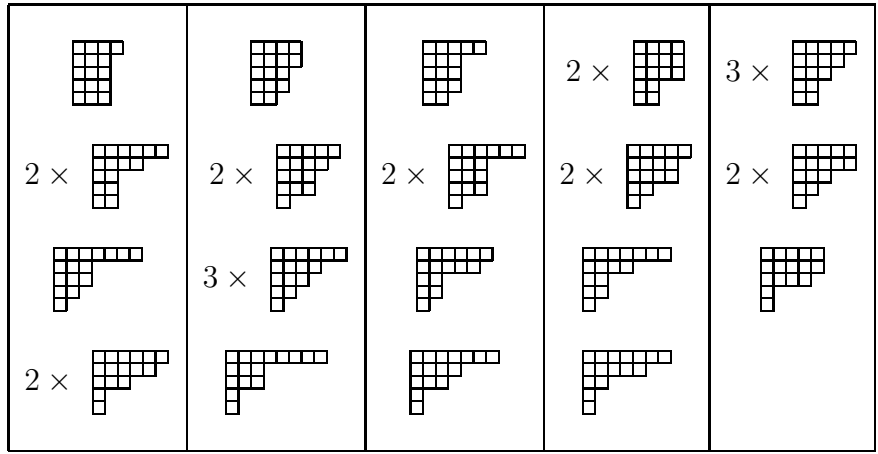
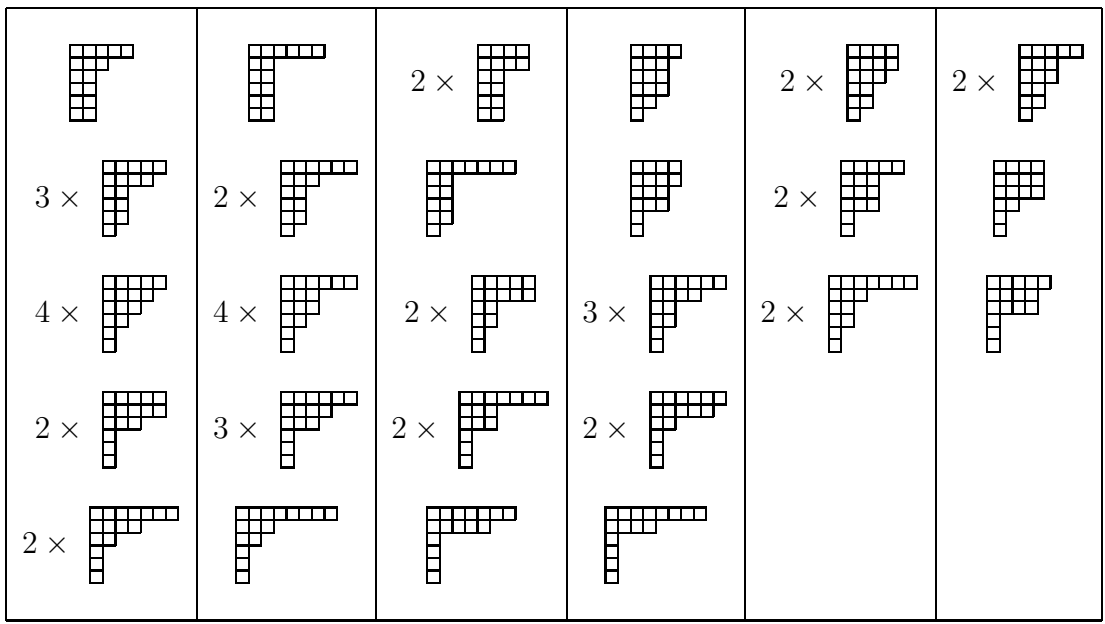
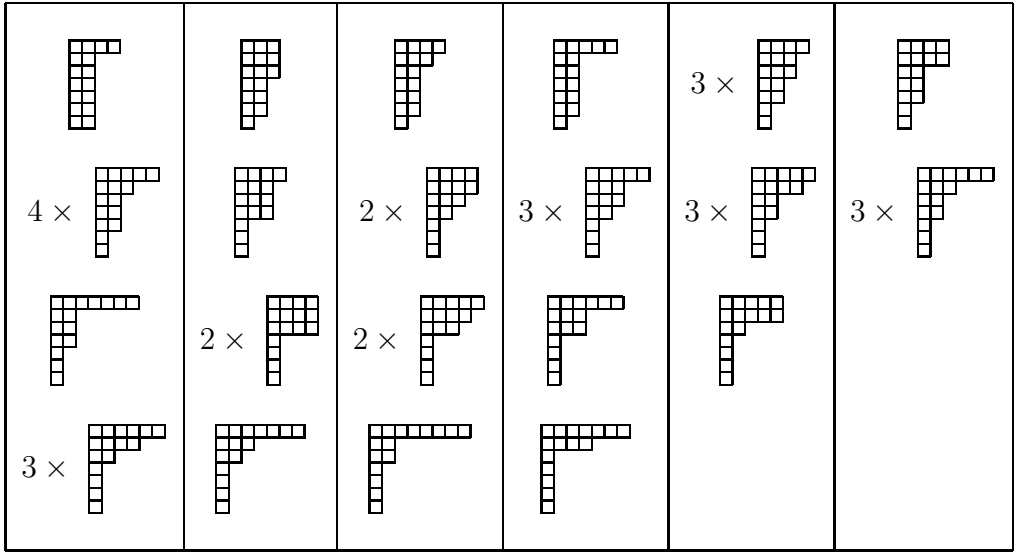
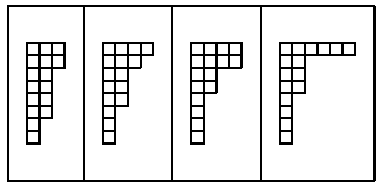
\renewcommand{\arraystretch}{0.8}
\begin{center}
\begin{tabular}{|c|c|c|c|c|}
\hline
&&&&\\
$\TT{6 6 3 1}$ &
$\TT{6 4 4 2}$ &
$\TT{6 4 3 3}$ &
$\TT{7 4 3 2}$ &
$\TT{7 3 3 3}$ \\
&&&&\\
\hline
\end{tabular}
\end{center}
\begin{center}
\begin{tabular}{|c|c|c|c|c|}
\hline
&&&&\\
$\TT{4 3 3 3 3}$ &
$\TT{4 4 3 3 2}$ &
$\TT{5 3 3 3 2}$ &
$2\times\ \TT{4 4 4 2 2}$ &
$3\times\ \TT{5 4 3 2 2}$ \\
&&&&\\
$2\times\ \TT{6 4 2 2 2}$ &
$2\times\ \TT{5 4 3 3 1}$ &
$2\times\ \TT{6 3 3 3 1}$ &
$2\times\ \TT{5 4 4 2 1}$ &
$2\times\ \TT{5 5 3 2 1}$ \\
&&&&\\
$\TT{7 3 3 2 1}$ &
$3\times\ \TT{6 4 3 2 1}$ &
$\TT{6 5 2 2 1}$ &
$\TT{7 4 2 2 1}$ &
$\TT{5 5 4 1 1}$ \\
&&&&\\
$2\times\ \TT{6 5 3 1 1}$ &
$\TT{8 3 3 1 1}$ &
$\TT{7 4 3 1 1}$ &
$\TT{7 5 2 1 1}$ & \\
&&&&\\
\hline
\end{tabular}
\end{center}
\begin{center}
\begin{tabular}{|c|c|c|c|c|c|}
\hline
&&&&&\\
$\TT{5 3 2 2 2 2}$ &
$\TT{6 2 2 2 2 2}$ &
$2\times\ \TT{4 4 2 2 2 2}$ &
$\TT{4 3 3 3 2 1}$ &
$2\times\ \TT{4 4 3 2 2 1}$ &
$2\times\ \TT{5 3 3 2 2 1}$ \\
&&&&&\\
$3\times\ \TT{5 4 2 2 2 1}$ &
$2\times\ \TT{6 3 2 2 2 1}$ &
$\TT{7 2 2 2 2 1}$ &
$\TT{4 4 3 3 1 1}$ &
$2\times\ \TT{5 3 3 3 1 1}$ &
$\TT{4 4 4 2 1 1}$ \\
&&&&&\\
$4\times\ \TT{5 4 3 2 1 1}$ &
$4\times\ \TT{6 3 3 2 1 1}$ &
$2\times\ \TT{5 5 2 2 1 1}$ &
$3\times\ \TT{6 4 2 2 1 1}$ &
$2\times\ \TT{7 3 2 2 1 1}$ &
$\TT{5 4 4 1 1 1}$ \\
&&&&&\\
$2\times\ \TT{5 5 3 1 1 1}$ &
$3\times\ \TT{6 4 3 1 1 1}$ &
$2\times\ \TT{7 3 3 1 1 1}$ &
$2\times\ \TT{6 5 2 1 1 1}$ & &\\
&&&&&\\
$2\times\ \TT{7 4 2 1 1 1}$ &
$\TT{8 3 2 1 1 1}$ &
$\TT{7 5 1 1 1 1}$ &
$\TT{8 4 1 1 1 1}$ && \\
&&&&&\\
\hline
\end{tabular}
\end{center}
\begin{center}
\begin{tabular}{|c|c|c|c|c|c|}
\hline
&&&&&\\
$\TT{4 2 2 2 2 2 2}$ &
$\TT{3 3 3 2 2 2 1}$ &
$\TT{4 3 2 2 2 2 1}$ &
$\TT{5 2 2 2 2 2 1}$ &
$3\times\ \TT{4 3 3 2 2 1 1}$ &
$\TT{4 4 2 2 2 1 1}$ \\
&&&&&\\
$4\times\ \TT{5 3 2 2 2 1 1}$ &
$\TT{4 3 3 3 1 1 1}$ &
$2\times\ \TT{4 4 3 2 1 1 1}$ &
$3\times\ \TT{5 3 3 2 1 1 1}$ &
$3\times\ \TT{5 4 2 2 1 1 1}$ &
$3\times\ \TT{6 3 2 2 1 1 1}$ \\
&&&&&\\
$\TT{7 2 2 2 1 1 1}$ &
$2\times\ \TT{4 4 4 1 1 1 1}$ &
$2\times\ \TT{5 4 3 1 1 1 1}$ &
$\TT{6 3 3 1 1 1 1}$ &
$\TT{5 5 2 1 1 1 1}$ & \\
&&&&&\\
$3\times\ \TT{6 4 2 1 1 1 1}$ &
$\TT{7 3 2 1 1 1 1}$ &
$\TT{8 2 2 1 1 1 1}$ &
$\TT{7 4 1 1 1 1 1}$ & &\\ 
&&&&&\\
\hline
\end{tabular}
\end{center}
\begin{center}
\begin{tabular}{|c|c|c|c|}
\hline
&&&\\
$\TT{3 3 2 2 2 2 1 1}$ &
$\TT{4 3 2 2 2 1 1 1}$ &
$\TT{4 4 2 2 1 1 1 1}$ &
$\TT{6 2 2 2 1 1 1 1}$ \\
&&&\\
\hline
\end{tabular}
\end{center}
We now find $569$ fundamental invariants of degree 18:~
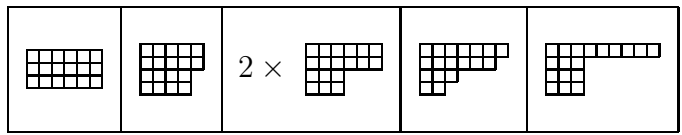
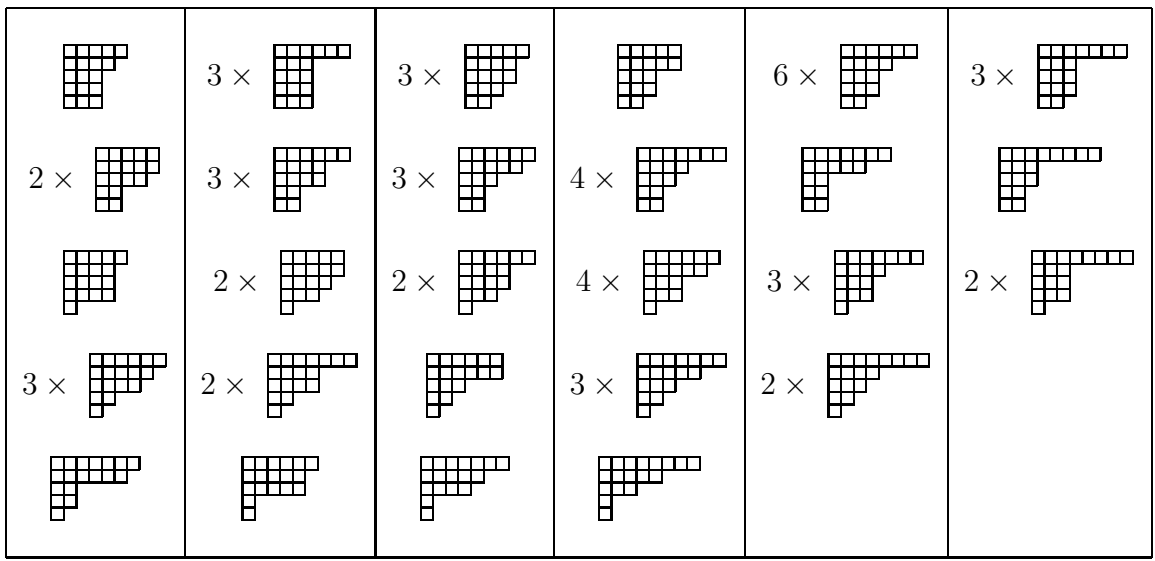
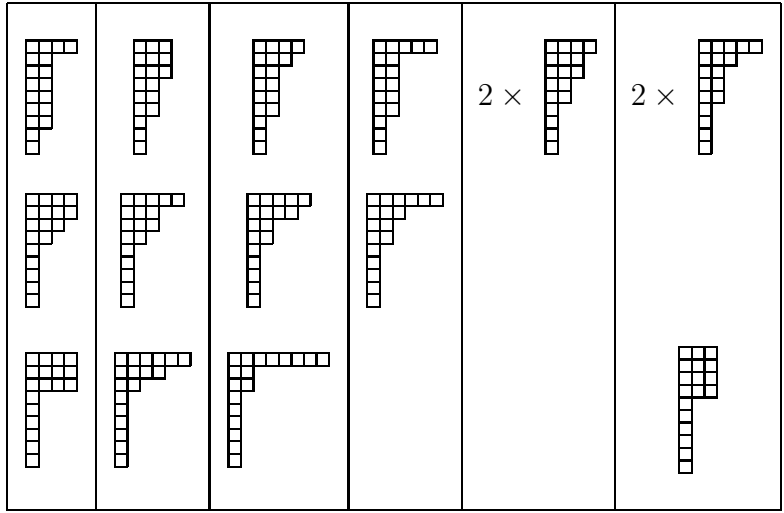
\begin{center}
\begin{tabular}{|c|c|c|c|c|}
\hline
&&&&\\
$\TT{6 6 6}$ &
$\TT{5 5 4 4}$ &
$2\times\ \TT{6 6 3 3}$ &
$\TT{7 6 3 2}$ &
$\TT{9 3 3 3}$ \\
&&&&\\
\hline
\end{tabular}
\end{center}
\begin{center}
\begin{tabular}{|c|c|c|c|c|c|}
\hline
&&&&&\\
$\TT{5 4 3 3 3}$ &
$3\times\ \TT{6 3 3 3 3}$ &
$3\times\ \TT{5 4 4 3 2}$ &
$\TT{5 5 3 3 2}$ &
$6\times\ \TT{6 4 3 3 2}$ &
$3\times\ \TT{7 3 3 3 2}$ \\
&&&&&\\
$2\times\ \TT{5 5 4 2 2}$ &
$3\times\ \TT{6 4 4 2 2}$ &
$3\times\ \TT{6 5 3 2 2}$ &
$4\times\ \TT{7 4 3 2 2}$ &
$\TT{7 5 2 2 2}$ &
$\TT{8 3 3 2 2}$ \\
&&&&&\\
$\TT{5 4 4 4 1}$ &
$2\times\ \TT{5 5 4 3 1}$ &
$2\times\ \TT{6 4 4 3 1}$ &
$4\times\ \TT{6 5 3 3 1}$ &
$3\times\ \TT{7 4 3 3 1}$ &
$2\times\ \TT{8 3 3 3 1}$ \\
&&&&&\\
$3\times\ \TT{6 5 4 2 1}$ &
$2\times\ \TT{7 4 4 2 1}$ &
$\TT{6 6 3 2 1}$ &
$3\times\ \TT{7 5 3 2 1}$ &
$2\times\ \TT{8 4 3 2 1}$ &\\
&&&&&\\
$\TT{7 6 2 2 1}$ &
$\TT{6 5 5 1 1}$ &
$\TT{7 5 4 1 1}$ &
$\TT{8 5 3 1 1}$ &&\\
&&&&&\\
\hline
\end{tabular}
\end{center}
\begin{center}
\begin{tabular}{|c|c|c|c|c|c|}
\hline
&&&&&\\
$\TT{3 3 3 3 3 3}$ &
$\TT{4 3 3 3 3 2}$ &
$5\times\ \TT{4 4 3 3 2 2}$ &
$2\times\ \TT{5 3 3 3 2 2}$ &
$2\times\ \TT{4 4 4 2 2 2}$ &
$6\times\ \TT{5 4 3 2 2 2}$ \\
&&&&&\\
$4\times\ \TT{6 3 3 2 2 2}$ &
$4\times\ \TT{5 5 2 2 2 2}$ &
$2\times\ \TT{6 4 2 2 2 2}$ &
$\TT{7 3 2 2 2 2}$ &
$\TT{4 4 3 3 3 1}$ &
$3\times\ \TT{5 3 3 3 3 1}$ \\
&&&&&\\
$3\times\ \TT{4 4 4 3 2 1}$ &
$10\times\ \TT{5 4 3 3 2 1}$ &
$5\times\ \TT{6 3 3 3 2 1}$ &
$7\times\ \TT{5 4 4 2 2 1}$ &
$6\times\ \TT{5 5 3 2 2 1}$ &
$12\times\ \TT{6 4 3 2 2 1}$ \\
&&&&&\\
$4\times\ \TT{7 3 3 2 2 1}$ &
$4\times\ \TT{6 5 2 2 2 1}$ &
$4\times\ \TT{7 4 2 2 2 1}$ &
$\TT{8 3 2 2 2 1}$ &
$3\times\ \TT{4 4 4 4 1 1}$ &
$5\times\ \TT{5 4 4 3 1 1}$ \\
&&&&&\\
\hline
\end{tabular}
\end{center}
\begin{center}
\begin{tabular}{|c|c|c|c|c|c|}
\hline
&&&&&\\
$7\times\ \TT{5 5 3 3 1 1}$ &
$6\times\ \TT{6 4 3 3 1 1}$ &
$4\times\ \TT{7 3 3 3 1 1}$ &
$5\times\ \TT{5 5 4 2 1 1}$ &
$8\times\ \TT{6 4 4 2 1 1}$ &
$\scriptstyle{10\times\ } \TT{6 5 3 2 1 1}$ \\
&&&&&\\
$8\times\ \TT{7 4 3 2 1 1}$ &
$2\times\ \TT{8 3 3 2 1 1}$ &
$4\times\ \TT{6 6 2 2 1 1}$ &
$3\times\ \TT{7 5 2 2 1 1}$ &
$3\times\ \TT{8 4 2 2 1 1}$ &
$2\times\ \TT{5 5 5 1 1 1}$ \\
&&&&&\\
$\TT{6 6 3 1 1 1}$ &
$3\times\ \TT{6 5 4 1 1 1}$ &
$3\times\ \TT{7 4 4 1 1 1}$ &
$5\times\ \TT{7 5 3 1 1 1}$ &
$2\times\ \TT{8 4 3 1 1 1}$ &\\
&&&&&\\
$\TT{9 3 3 1 1 1}$ &
$\TT{7 6 2 1 1 1}$ &
$2\times\ \TT{8 5 2 1 1 1}$ &
$\TT{9 4 2 1 1 1}$ &
$\TT{7 7 1 1 1 1}$ &\\
&&&&&\\
\hline
\end{tabular}
\end{center}
\begin{center}
\begin{tabular}{|c|c|c|c|c|c|}
\hline
&&&&&\\
$2\times\ \TT{4 3 3 2 2 2 2}$ &
$2\times\ \TT{5 3 2 2 2 2 2}$ &
$\TT{6 2 2 2 2 2 2}$ &
$2\times\ \TT{4 3 3 3 2 2 1}$ &
$6\times\ \TT{5 3 3 2 2 2 1}$ &
$5\times\ \TT{4 4 3 2 2 2 1}$ \\
&&&&&\\
$5\times\ \TT{5 4 2 2 2 2 1}$ &
$4\times\ \TT{6 3 2 2 2 2 1}$ &
$\TT{7 2 2 2 2 2 1}$ &
$\TT{4 3 3 3 3 1 1}$ &
$4\times\ \TT{4 4 3 3 2 1 1}$ &
$6\times\ \TT{5 3 3 3 2 1 1}$ \\
&&&&&\\
$4\times\ \TT{4 4 4 2 2 1 1}$ &
$13\times\ \TT{5 4 3 2 2 1 1}$ &
$9\times\ \TT{6 3 3 2 2 1 1}$ &
$4\times\ \TT{5 5 2 2 2 1 1}$ &
$10\times\ \TT{6 4 2 2 2 1 1}$ &
$5\times\ \TT{7 3 2 2 2 1 1}$ \\
&&&&&\\
\hline
\end{tabular}
\end{center}
\begin{center}
\begin{tabular}{|c|c|c|c|c|c|}
\hline
&&&&&\\
$\TT{8 2 2 2 2 1 1}$ &
$3\times\ \TT{4 4 4 3 1 1 1}$ &
$7\times\ \TT{5 4 3 3 1 1 1}$ &
$5\times\ \TT{6 3 3 3 1 1 1}$ &
$7\times\ \TT{5 4 4 2 1 1 1}$ &
$8\times\ \TT{5 5 3 2 1 1 1}$ \\
&&&&&\\
$14\times\ \TT{6 4 3 2 1 1 1}$ &
$7\times\ \TT{7 3 3 2 1 1 1}$ &
$8\times\ \TT{6 5 2 2 1 1 1}$ &
$8\times\ \TT{7 4 2 2 1 1 1}$ &
$4\times\ \TT{8 3 2 2 1 1 1}$ &
$\TT{9 2 2 2 1 1 1}$ \\
&&&&&\\
$3\times\ \TT{5 5 4 1 1 1 1}$ &
$5\times\ \TT{6 4 4 1 1 1 1}$ &
$6\times\ \TT{6 5 3 1 1 1 1}$ &
$7\times\ \TT{7 4 3 1 1 1 1}$ &
$2\times\ \TT{8 3 3 1 1 1 1}$ &
$2\times\ \TT{6 6 2 1 1 1 1}$ \\
&&&&&\\
$5\times\ \TT{7 5 2 1 1 1 1}$ &
$4\times\ \TT{8 4 2 1 1 1 1}$ &
$\TT{9 3 2 1 1 1 1}$ &
$\TT{7 6 1 1 1 1 1}$ &
$\TT{8 5 1 1 1 1 1}$ &
$\TT{9 4 1 1 1 1 1}$ \\
&&&&&\\
\hline
\end{tabular}
\end{center}
\begin{center}
\begin{tabular}{|c|c|c|c|c|c|}
\hline
&&&&&\\
$\TT{3 3 2 2 2 2 2 2}$ &
$2\times\ \TT{4 3 2 2 2 2 2 1}$ &
$\TT{5 2 2 2 2 2 2 1}$ &
$2\times\ \TT{3 3 3 3 2 2 1 1}$ &
$4\times\ \TT{4 3 3 2 2 2 1 1}$ &
$4\times\ \TT{4 4 2 2 2 2 1 1}$ \\
&&&&&\\
$4\times\ \TT{5 3 2 2 2 2 1 1}$ &
$2\times\ \TT{6 2 2 2 2 2 1 1}$ &
$3\times\ \TT{4 3 3 3 2 1 1 1}$ &
$5\times\ \TT{4 4 3 2 2 1 1 1}$ &
$9\times\ \TT{5 3 3 2 2 1 1 1}$ &
$6\times\ \TT{5 4 2 2 2 1 1 1}$ \\
&&&&&\\
$6\times\ \TT{6 3 2 2 2 1 1 1}$ &
$\TT{7 2 2 2 2 1 1 1}$ &
$5\times\ \TT{4 4 3 3 1 1 1 1}$ &
$\TT{5 3 3 3 1 1 1 1}$ &
$\TT{4 4 4 2 1 1 1 1}$ &
$10\times\ \TT{5 4 3 2 1 1 1 1}$ \\
&&&&&\\
\hline
\end{tabular}
\end{center}
\begin{center}
\begin{tabular}{|c|c|c|c|c|c|}
\hline
&&&&&\\
$6\times\ \TT{6 3 3 2 1 1 1 1}$ &
$5\times\ \TT{5 5 2 2 1 1 1 1}$ &
$6\times\ \TT{6 4 2 2 1 1 1 1}$ &
$6\times\ \TT{7 3 2 2 1 1 1 1}$ &
$\TT{5 5 3 1 1 1 1 1}$ &
$3\times\ \TT{5 4 4 1 1 1 1 1}$ \\
&&&&&\\
$6\times\ \TT{6 4 3 1 1 1 1 1}$ &
$\TT{7 3 3 1 1 1 1 1}$ &
$3\times\ \TT{6 5 2 1 1 1 1 1}$ &
$4\times\ \TT{7 4 2 1 1 1 1 1}$ &&\\
&&&&&\\
$2\times\ \TT{8 3 2 1 1 1 1 1}$ &
$\TT{9 2 2 1 1 1 1 1}$ &
$2\times\ \TT{6 6 1 1 1 1 1 1}$ & 
$\TT{8 4 1 1 1 1 1 1}$ &&\\
&&&&&\\
\hline
\end{tabular}
\end{center}
\begin{center}
\begin{tabular}{|c|c|c|c|c|c|}
\hline
&&&&&\\
$\TT{4 2 2 2 2 2 2 1 1}$ &
$\TT{3 3 3 2 2 2 1 1 1}$ &
$\TT{4 3 2 2 2 2 1 1 1}$ &
$\TT{5 2 2 2 2 2 1 1 1}$ &
$2\times\ \TT{4 3 3 2 2 1 1 1 1}$ &
$2\times\ \TT{5 3 2 2 2 1 1 1 1}$ \\
&&&&&\\
$\TT{4 4 3 2 1 1 1 1 1}$ &
$\TT{5 3 3 2 1 1 1 1 1}$ &
$\TT{5 4 2 2 1 1 1 1 1}$ &
$\TT{6 3 2 2 1 1 1 1 1}$ &&\\
&&&&&\\
$\TT{4 4 4 1 1 1 1 1 1}$ &
$\TT{6 4 2 1 1 1 1 1 1}$ &
$\TT{8 2 2 1 1 1 1 1 1}$ &&&
$\TT{3 3 3 3 1 1 1 1 1 1}$\\
&&&&&\\
\hline
\end{tabular}
\end{center}
So far, we have identified 740 basic invariants, up to degree 18. 
Since $E_7$ has an $SO(12)$ subgroup, there is 
at least a $({\bf 12})^{12}$ invariant along this flat direction. 
We found only one invariant with as many as $10$ antisymmetrized
boxes thus far. This indicates that many more invariants are yet 
to be found.

\subsection{Status of the self-dual model for $E_7$ with 4 flavors}

In \cite{Cho:1998am}, Cho studied in detail the flat directions 
of the $E_7$ self-dual model of \cite{PouliotStrassler,Distler:1997ub}
and found some problems in matching invariants. As for $E_6$ with 
$6$ flavors, we find that invariants had been missed; while some of
the new invariants solve the problems mentioned in~\cite{Cho:1998am}, 
others cause new problems.
The list of invariants for $E_7$ with 4 flavors have the following 
$SU(4)\times U(1)_R$ charge:
\renewcommand{\arraystretch}{1.2}
\begin{center}
\begin{tabular}{|c|c|c|}
\hline
($\TT{1 1}$, $1/2$) & ($\TT{4}$, $1$) & ($\TT{3 3}$, $3/2$) \\
($\TT{4 2 2}$, $2$) &
($\TT{1 1}$, $5/2$) &
($\TT{4 1 1}$, $5/2$) \\
($\TT{3 3}$, $5/2$) &
($\TT{4 4 4}$, $3$) &
($\TT{4 3 1}$, $3$) \\
($\TT{4}$, $3$) &
($\TT{3 2 1}$, $7/2$) &
($\TT{5 3 2}$, $7/2$) \\
($\TT{3 3}$, $7/2$) &
($\TT{1 1}$, $7/2$) &
($\TT{5 5 2}$, $4$) \\
($\TT{4 2 2}$, $4$) &
($\TT{3 1}$, $4$) &
($\TT{5 2 1}$, $4$) \\
($\TT{4}$, $4$) &
($\TT{6 6 6}$, $9/2$) &
($\TT{1 1}$, $9/2$) \\
($\TT{3 3}$, $9/2$) &
($\TT{5 4 1}$, $9/2$) &
($\TT{6}$, $9/2$) \\
\hline
\end{tabular}
\end{center}
A few representations in this table are complex and are then 
potentially dangerous. A much more detailed analysis
would be required to argue either way whether this duality is valid or not
\footnote{Just as for the $E_6$ self-dual model, there is a serious
problem with matching the global discrete symmetry. 
One way around this problem is
to add the invariant $I_4$ to the superpotential of both electric
and magnetic theories. This reduces the global $SU(4)$ symmetry to $SU(3)$.}.

\section{Examples with Other Groups}
We have chosen a few other examples that illustrate the complexity 
that one encounters 
for some familiar rings that arise for supersymmetric gauge theories.
\subsection{The ring of $SU(2)$ with fundamentals}
In this section, we describe the simplest example of all in greater detail. 
We construct explicitly the free resolution of the rings, again up 
to degree 18.
The moduli space of $SU(2)$ with $2N$ doublets 
is described by one flat direction. Alternatively,
it can be described in a gauge invariant way by the 
invariant~\cite{Seiberg:1994bz}:
\begin{eqnarray}
V_{ij}=\epsilon_{\alpha\beta} Q^\alpha_i Q^\beta_j= \ \Ytwo \nonumber\ .
\end{eqnarray}
This statement is known as the ``first fundamental theorem of 
classical invariant theory,''
and has of course appropriate generalizations for the $SU$, 
$SO$ and $Sp$ groups.
This invariant satisfies one single constraint (one first-order syzygy):
\begin{eqnarray}
Z_1 = \textrm{Pf\ } V = \Yfour\nonumber \ .
\end{eqnarray}
That statement is known as the ``second fundamental theorem.''
This constraint is itself constrained (one second-order syzygy):
\begin{eqnarray}
Z_2 =\ \Xfive\Xone \nonumber \ .
\end{eqnarray}
The second-order syzygy is constrained by 2 third-order syzygies:
\begin{eqnarray}
Z_3^1=\ \Xsix\Xone\Xone\qquad Z_3^2=\ \Xfive\Xfive \nonumber \ .
\end{eqnarray}
The third-order syzygies are constrained by 2 fourth-order syzygies:
\begin{eqnarray}
Z_4^1=\ \Xseven\Xone\Xone\Xone\qquad Z_4^2=\ \Xsix\Xfive\Xone\nonumber \ .
\end{eqnarray}
The fourth-order syzygies are constrained by 3 fifth-order syzygies:
\begin{eqnarray}
Z_5^1=\ \Xn8\Xone\Xone\Xone\Xone\qquad Z_5^2=\ \Xsix\Xsix\Xtwo\qquad 
Z_5^3=\ \Xseven\Xfive\Xone\Xone\nonumber\ .
\end{eqnarray}
The fifth-order syzygies are constrained by 4 sixth-order syzygies:
\begin{eqnarray}
Z_6^1&=&\  \Xsix\Xsix\Xsix\qquad 
Z_6^2=\ \Xseven\Xsix\Xtwo\Xone \nonumber\\
Z_6^3&=&\ \Xn8\Xfive\Xone\Xone\Xone\qquad 
Z_6^4=\ \Xn9\Xone\Xone\Xone\Xone\Xone\nonumber\ .
\end{eqnarray}
And the sixth-order syzygies are constrained by 4 seventh-order syzygies:
\begin{eqnarray}
Z_7^1&=&\ \Xn{10}\Xone\Xone\Xone\Xone\Xone\Xone \qquad
Z_7^2=\ \Xseven\Xseven\Xtwo\Xtwo \nonumber\\
Z_7^3&=&\ \Xn8\Xsix\Xtwo\Xtwo \qquad
Z_7^4=\ \Xn9\Xfive\Xone\Xone\Xone\Xone \nonumber \ .
\end{eqnarray}
Then the seventh-order syzygies are constrained by at least one eight-order syzygies:
\begin{eqnarray}
Z_8=\Xn{11}\Xone\Xone\Xone\Xone\Xone\Xone\Xone\nonumber \ ,
\end{eqnarray}   
and so on and so forth. Exhibiting this syzygy chain is known as 
constructing the free resolution of the ring.
To our knowledge, the detailed form of higher-order syzygies has 
not played a role
in physics.

\subsection{Invariants and syzygies of $G_2$ with fundamentals}
It is only since the eighties~\cite{schwarz:1983,Schwarz:1988} 
that the first and second fundamental theorems have been
proven for $G_2$, and of course the other exceptional groups are
well beyond reach.
In this section, we will verify explicitly the results 
of~\cite{schwarz:1983,Schwarz:1988} for the invariants and the 
first-order syzygies using our computer-based method.
We find a complete agreement, and we extend the results to the 
second-order syzygies.
For the 7-dimensional fundamental representation, the invariants 
are well-known to be: 
\begin{eqnarray}
I_1=\rseven^2_{[2]} =\  \Yone\Yone \qquad I_2=\rseven^3_{[1^3]}=\ 
\Ythree \qquad I_3=\rseven^4_{[1^4]} =\ \Yfour\nonumber\ .
\end{eqnarray}
There are first-order constraints among these invariants:
\begin{eqnarray}
C_{(6)} &=& \left( I_1 I_3 + I_2^2 = 0 \right)_{[1^5\, 1]} =\  
\Xfive\Xone \qquad
C^{1}_{(7)} = \left( I_2 I_3 = 0 \right)_{[1^6\, 1]} =\  
\Xsix\Xone \nonumber\  \\
C^{2}_{(7)} &=& \left( I_1^2 I_2 + I_2 I_3 = 0 
\right)_{[1^5\, 1^2]} =\  \Xfive\Xtwo \qquad
C^{1}_{(8)} = \left( I_3^2 = 0 \right)_{[1^8]}=\ 
\Xn8 \nonumber\ \\
C^{2}_{(8)} &=& \left( I_1^4 +I_3^2 + I_1I_2^2 = 
0\right)_{[2^4]} =\  \Yfour\Yfour \qquad
C^{3}_{(8)} = \left( I_1^2 I_3 + \alpha\ I_1 I_2^2 + 
I_3^2 = 0\right)_{[1^6\, 1^2]} =\  \Xsix\Xtwo \nonumber\ . 
\end{eqnarray}
The subscript for $C$ denotes the degree of the constraint, 
while the superscript just enumerates them.  
It is claimed in \cite{Schwarz:1988} that these are all the 
constraints, and our explicit calculation checks this to degree 16.
The information in~\cite{Schwarz:1988} is more detailed than 
what our computer-based method lets us achieve.
For example, we see, from table~I in \cite{Schwarz:1988}, that 
the coefficient $\alpha$ is zero in the constraint $C^{3}_{(8)}$.
Note also that our constraints are listed as $(5.4.1)$ through 
$(5.4.5)$ of table I. The other constraints in table~I, as
shown in~\cite{Schwarz:1988}, follow from the constraints 
$(5.4.1)$ through $(5.4.5)$, and therefore should not be included
as independent constraints; our calculation explicitly confirms 
that. We now move on beyond the results of \cite{Schwarz:1988}.  
In degree 9, we find two second-order syzygies, i.e.  constraints 
amongst the first-order constraints:
\begin{eqnarray}
Z_{(9)}^1=\left( I_1 C_{(7)}^1 + I_1 C_{(7)}^2 + I_2 C_{(6)} = 0 
\right)_{[1^6\, 1^2\, 1]}=\ \Xsix\Xtwo\Xone \nonumber\ \\
Z_{(9)}^2=\left( I_1 C_{(7)}^1 + I_2 C_{(6)} = 0 
\right)_{[1^7\, 1^2]}=\ \Xseven\Xtwo\nonumber\ .
\end{eqnarray}
In degree 10, we find twelve second-order syzygies:
\begin{eqnarray}
Z_{(10)}^1 = \left(  I_1 C_{(8)}^2 + I_2  C_{(7)}^2 + 
I_3 C_{(6)} = 0 \right)_{[1^5\,1^4\,1]} =\  \Xfive\Xfour\Xone \nonumber\\
Z_{(10)}^2 = \left(  I_1 C_{(8)}^3 + I_2 C_{(7)}^1 + 
I_2 C_{(7)}^2+ I_1^2 C_{(6)}+ I_3 C_{(6)} = 0 
\right)_{[1^6\,1^3\,1]} =\ \Xsix\Xthree\Xone \nonumber\\
Z_{(10)}^{3,4} = \left(  I_2 C_{(7)}^1 +  
I_2 C_{(7)}^2 + I_3 C_{(6)} = 0 \right)_{[1^6\,1^4]} =\ 
\Xsix\Xfour \nonumber\\
Z_{(10)}^5 = \left( I_2  C_{(7)}^2 + I_1^2 C_{(6)}  = 0 
\right)_{[1^6\,2^2]} =\ \Xsix\Xtwo\Xtwo \nonumber\\
Z_{(10)}^6 = \left( I_1  C_{(8)}^3 + I_2 C_{(7)}^1 +  
I_2C_{(7)}^2+ I_3 C_{(6)}+ I_1^2 C_{(6)} = 0 
\right)_{[1^7\,1^3]} =\  \Xseven\Xthree \nonumber\\
Z_{(10)}^{7,8} = \left(  I_1  C_{(8)}^3 +  I_2 C_{(7)}^1 + 
I_2 C_{(7)}^2+ I_1^2 C_{(6)}+ I_3 C_{(6)} = 0 
\right)_{[1^7\,1^2\,1]} =\ \Xseven\Xtwo\Xone \nonumber\\
Z_{(10)}^{9,10} = \left(  I_2 C_{(7)}^1 + I_2 C_{(7)}^2 + 
I_3 C_{(6)} = 0 \right)_{[1^8\,1^2]} =\  \Xn8\Xtwo \nonumber\\
Z_{(10)}^{11} = \left(  I_1 C_{(8)}^1 + I_2 C_{(7)}^1 + 
I_3 C_{(6)} = 0 \right)_{[1^8\,2]} =\ \Xn8\Xone\Xone \nonumber\\
Z_{(10)}^{12} = \left(  I_1 C_{(8)}^1 + I_2 C_{(7)}^1 + 
I_3 C_{(6)} = 0 \right)_{[1^9\,1]} =\ \Xn9\Xone \nonumber\ .
\end{eqnarray}
In degree 11, there are twenty-one second-order syzygies:
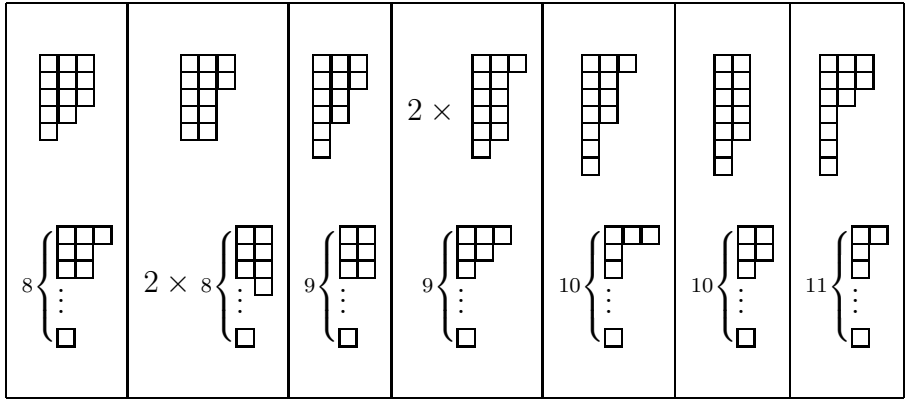
\begin{center}
\begin{tabular}{|c|c|c|c|c|}
\hline
&&&&\\
$\Xfive\Xfive\Xone$
&$\Xfive\Xfour\Xtwo$ 
&$2\times\ \Xsix\Xfive$
&$2\times\ \Xsix\Xfour\Xone$
&$\Xsix\Xthree\Xtwo$ \\
&&&&\\
$2\times\ \Xseven\Xthree\Xone$
&$2\times\ \Xseven\Xfour$ 
&$\Xseven\Xtwo\Xtwo$
&$2\times\ \Xn8\Xtwo\Xone$
&$2\times\ \Xn8\Xthree$ \\
&&&&\\
$2\times\ \Xn9\Xtwo$
&$2\times\ \Xn9\Xone\Xone$   
&$\Xn{10}\Xone$ &&\\
&&&&\\
\hline
\end{tabular}
\end{center}
In degree 12, we find the first third-order syzygies:
\begin{eqnarray}
\Xseven\Xtwo\Xtwo\Xone\qquad  \Xn8\Xtwo\Xtwo \nonumber\ ,
\end{eqnarray}
as well as sixteen more second-order syzygies:
\begin{center}
\begin{tabular}{|c|c|c|c|c|c|c|}
\hline
&&&&&&\\
$\Xfive\Xfour\Xthree$
&$\Xfive\Xfive\Xtwo$
&$\Xsix\Xfour\Xtwo$
&$2\times\ \Xsix\Xfive\Xone$
&$\Xseven\Xfour\Xone$ 
&$\Xseven\Xfive$
&$\Xseven\Xthree\Xtwo$\\
&&&&&&\\
$\Xn8\Xthree\Xone$
&$2\times\ \Xn8\Xfour$
&$\Xn9\Xthree$ 
&$\Xn9\Xtwo\Xone$
&$\Xn{10}\Xone\Xone$ 
&$\Xn{10}\Xtwo$
&$\Xn{11}\Xone$  \\
&&&&&&\\
\hline
\end{tabular}
\end{center}    
In degree 13, there is only one second-order syzygy:
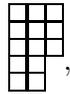
\begin{eqnarray}
\Xfive\Xfive\Xthree \nonumber,
\end{eqnarray}
but there are many third-order syzygies:
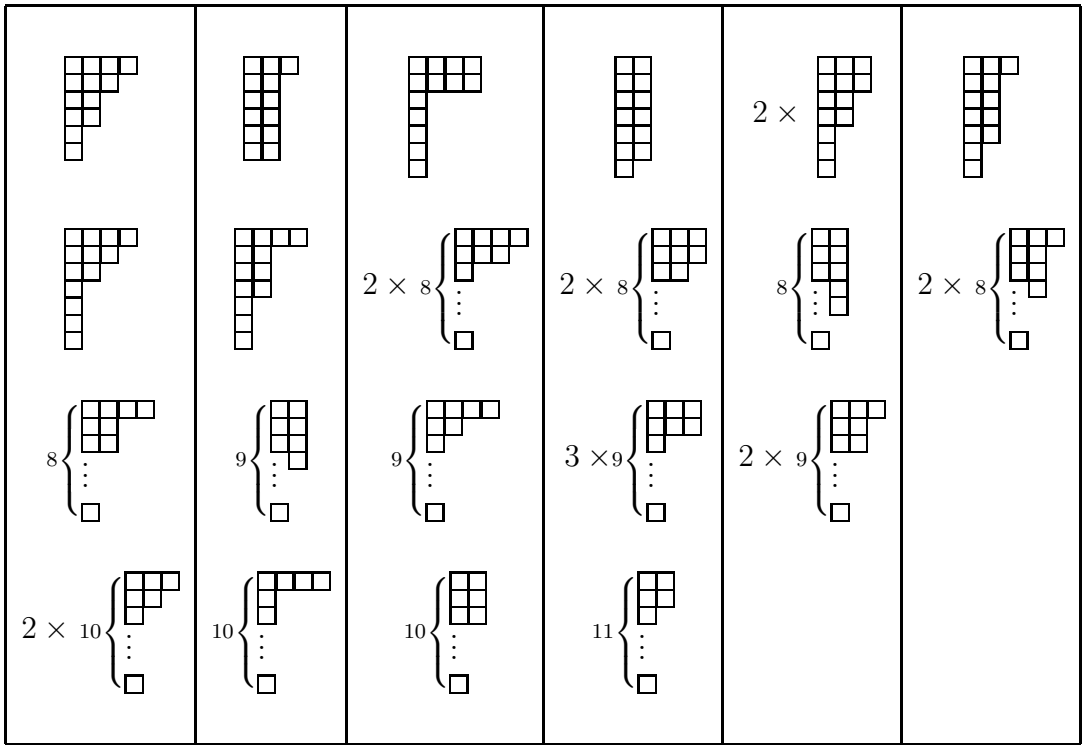
\begin{center}
\begin{tabular}{|c|c|c|c|c|c|}
\hline
&&&&&\\
$\Xsix\Xfour\Xtwo\Xone$
&$\Xsix\Xsix\Xone$ 
&$\Xseven\Xtwo\Xtwo\Xtwo$
&$\Xseven\Xsix$
&$2\times\ \Xseven\Xfour\Xtwo$
&$\Xseven\Xfive\Xone$\\
&&&&&\\
$\Xseven\Xthree\Xtwo\Xone$
&$\Xseven\Xfour\Xone\Xone$ 
&$2\times\ \Xn8\Xtwo\Xtwo\Xone$
&$2\times\ \Xn8\Xthree\Xtwo$ 
&$\Xn8\Xfive$
&$2\times\ \Xn8\Xfour\Xone$ \\
&&&&&\\
$\Xn8\Xthree\Xone\Xone$ 
&$\Xn9\Xfour$
&$\Xn9\Xtwo\Xone\Xone$
&$3\times \Xn9\Xtwo\Xtwo$
&$2\times\ \Xn9\Xthree\Xone$ &\\
&&&&&\\
$2\times\ \Xn{10}\Xtwo\Xone$ 
&$\Xn{10}\Xone\Xone\Xone$ 
&$\Xn{10}\Xthree$
&$\Xn{11}\Xtwo$  &&\\
&&&&&\\
\hline
\end{tabular}
\end{center}
In degree 14, 15 and 16, we found no new second-order syzygy 
(but hundreds of new third-order syzygies).

\subsection{Invariants of $SU(3)$ with symmetric tensors}
We repeat our search for the invariants of the $6$-dimensional 
symmetric tensor of $SU(3)$. We find this of interest because of 
the close resemblance, initially, between
the chiral ring of the $6$ of $SU(3)$ and that of the $27$ of $E_6$.
For the six-dimensional symmetric representation, the invariants are:
\begin{eqnarray}
I_1&=&\6^3_{[3]}=\ \Yone\Yone\Yone \qquad I_2=\6^6_{[2^3]}=\ 
\Ythree\Ythree \nonumber\\
I_3&=&Q^9_{[1^6\,2^2]}=\ \Xfive\Xtwo\Xtwo \qquad I_4=\6^6_{[1^6]}=\ 
\Xsix \nonumber\ .
\end{eqnarray}
These appears to be  all the invariants (at least up to degree 18). 
There is one constraint in degree 9:
\begin{eqnarray}
C_{(9)} = \left( I_1 I_3 = 0 \right)_{[1^7\,2]}=\  
\Xseven\Xone\Xone\nonumber\ .
\end{eqnarray}
In degree 12, we find seventeen of constraints:
\begin{center}
\begin{tabular}{|c|c|c|c|c|c|}
\hline
&&&&&\\
$\Xfour\Xfour\Xone\Xone\Xone\Xone$    
&$\Xfour\Xfour\Xtwo\Xtwo$
&$\Xfive\Xfive\Xone\Xone$    
&$\Xfive\Xthree\Xtwo\Xone\Xone$     
&$\Xfive\Xthree\Xthree\Xone$
&$\Xfive\Xfour\Xone\Xone\Xone$\\
&&&&&\\
$\Xsix\Xsix$
&$\Xsix\Xthree\Xthree$ 
&$\Xsix\Xfour\Xtwo$
&$\Xsix\Xthree\Xtwo\Xone$
&$\Xsix\Xtwo\Xtwo\Xone\Xone$&\\
&&&&&\\
$\Xn8\Xfour$
&$\Xn9\Xthree$
&$\Xseven\Xthree\Xtwo$
&$\Xn8\Xthree\Xone$ 
&$\Xn{10}\Xtwo$
&$\Xn{12}$\\
&&&&&\\
\hline
\end{tabular}
\end{center}  
As well as degree 12 second-order syzygies:
\begin{eqnarray}
\Xn8\Xone\Xone\Xone\Xone\qquad \Xn8\Xtwo\Xtwo\nonumber\ .
\end{eqnarray}
\subsection{Invariants of $SU(4)$ with symmetric tensors}
For $SU(4)$ with a $10$-dimensional symmetric tensor,
there is one  invariant of degree 4 and three of degree 8:
\begin{eqnarray}
I_1 &=& \rten^4_{[4]}=\ \Yone\Yone\Yone\Yone \qquad 
I_2 = \rten^8_{[2^4]}=\ \Yfour\Yfour \nonumber\\
I_3&=&\rten^8_{[1^6\, 2]}=\ \Xsix\Xone\Xone \qquad 
I_4=\rten^8_{[2^3\, 2]} =\  \Ythree\Ythree\Yone\Yone
\nonumber\ .
\end{eqnarray}
We then find a large number of invariants of degrees 12:
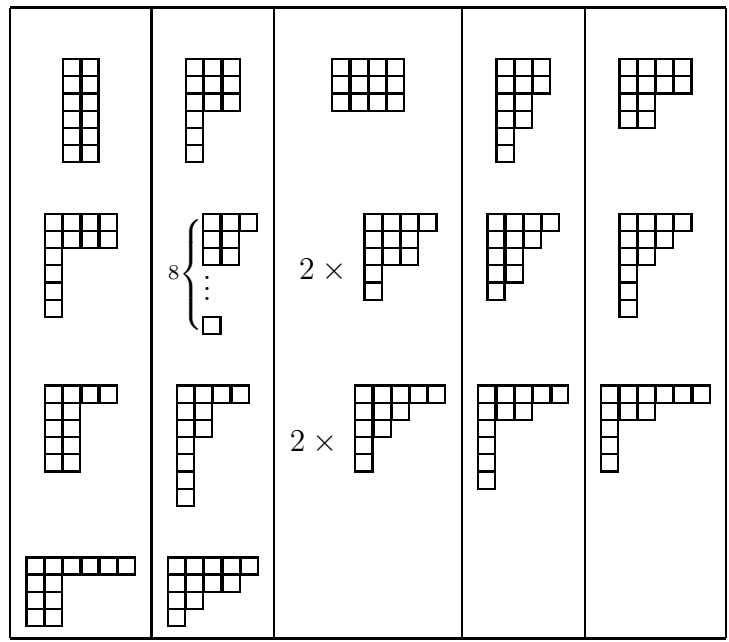
\begin{center}
\begin{tabular}{|c|c|c|c|c|}
\hline
&&&&\\
$\Xsix\Xsix$
&$\Xsix\Xthree\Xthree$
&$\Xthree\Xthree\Xthree\Xthree$
&$\Xsix\Xfour\Xtwo$
&$\Xfour\Xfour\Xtwo\Xtwo$ \\
&&&&\\
$\Xsix\Xtwo\Xtwo\Xtwo$
&$\Xn8\Xthree\Xone$
&$2\times\ \Xfive\Xthree\Xthree\Xone$
&$\Xfive\Xfour\Xtwo\Xone$
&$\Xsix\Xthree\Xtwo\Xone$ \\
&&&&\\
$\Xfive\Xfive\Xone\Xone$
&$\Xseven\Xthree\Xone\Xone$      
&$2\times\ \Xfive\Xthree\Xtwo\Xone\Xone$     
&$\Xsix\Xtwo\Xtwo\Xone\Xone$
&$\Xfive\Xtwo\Xtwo\Xone\Xone\Xone$ \\
&&&&\\
$\Xfour\Xfour\Xone\Xone\Xone\Xone$ 
&$\Xfour\Xthree\Xtwo\Xtwo\Xone$   &&&\\ 
\hline
\end{tabular}
\end{center}    
As well as one constraint of degree 12:
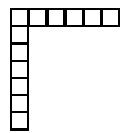
\begin{eqnarray}
\Xseven\Xone\Xone\Xone\Xone\Xone\nonumber \ .
\end{eqnarray}
There is an impressionistic resemblance between 
the chiral ring of symmetric
tensors of $SU(4)$ and the chiral ring of $E_7$ with 
fundamentals, at least for the invariants
of degrees 4, 8 and 12.
\subsection{Invariants of $SU5)$ with symmetric tensors}
For the $15$-dimensional symmetric tensor representation of 
$SU(5)$, the first few invariants are:
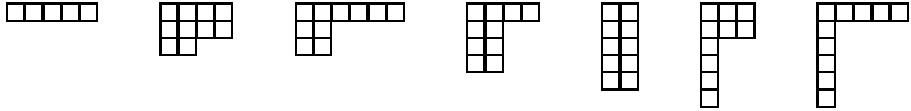
\begin{eqnarray}
\Xone\Xone\Xone\Xone\Xone \qquad \Xthree\Xthree\Xtwo\Xtwo 
\qquad \Xthree\Xthree\Xone\Xone\Xone\Xone\qquad 
\Xfour\Xfour\Xone\Xone \qquad \Xfive\Xfive\qquad \Xsix\Xtwo\Xtwo \qquad 
 \Xsix\Xone\Xone\Xone\Xone  \nonumber
\end{eqnarray}
Then in degree 15, there is roughly $178$ new invariants and one constraint.

\section{Conclusion}
In this paper, we have explored the structure of the chiral rings 
of several classes
of gauge theories. We were surprised that the invariants turned out 
to be so complicated.
This makes the search for more duals at least arduous, if not futile.

There might be a guide for finding more gauge theories that have a chance
of having a simple enough solution: one can systematically calculate the
homological dimension of rings of gauge theories and look
for examples, (e.g. by truncating the chiral ring~\cite{Kutasov:1995ve}),
which have a small homological dimension.

\acknowledgments

We would like to thank Dan Freed for explanations putting 
classical invariant theory in context.
This work was supported by NSF grant PHY-0071512.

\newpage

\appendix

\renewcommand{\arraystretch}{0.8}

\section{The constraints of degree 18 for $E_6$}
\begin{center}
\begin{tabular}{|c|c|c|c|c|c|}
\hline
&&&&&\\
$\Xfive\Xfour\Xfour\Xthree\Xtwo$
&$\Xfive\Xfive\Xfour\Xtwo\Xtwo$
&$\Xfive\Xfive\Xthree\Xthree\Xone\Xone$ &&&\\
&&&&&\\
$\Xsix\Xfive\Xfour\Xthree$
&$\Xsix\Xthree\Xthree\Xthree\Xthree$
&$\Xsix\Xfive\Xfive\Xtwo$
&$\Xsix\Xfive\Xthree\Xtwo\Xtwo$
&$\Xsix\Xfour\Xfour\Xthree\Xone$
&$\Xsix\Xfive\Xthree\Xthree\Xone$\\
&&&&&\\
$2\times\ \Xsix\Xfive\Xfour\Xtwo\Xone$
&$\Xsix\Xsix\Xthree\Xtwo\Xone$
&$\Xsix\Xfour\Xthree\Xtwo\Xtwo\Xone$
&$\Xsix\Xsix\Xfour\Xone\Xone$
&$\Xsix\Xfour\Xfour\Xtwo\Xone\Xone$
&$\Xsix\Xfive\Xthree\Xtwo\Xone\Xone$\\
&&&&&\\
$\Xseven\Xfour\Xthree\Xtwo\Xtwo$
&$\Xseven\Xseven\Xfour$ 
&$2\times\ \Xseven\Xfour\Xfour\Xthree$
&$3\times\ \Xseven\Xfive\Xfour\Xtwo$
&$\Xseven\Xsix\Xthree\Xtwo$
&$\Xseven\Xfive\Xtwo\Xtwo\Xtwo$\\
&&&&&\\
$\Xseven\Xthree\Xtwo\Xtwo\Xtwo\Xtwo$
&$2\times\ \Xseven\Xsix\Xfour\Xone$
&$\Xseven\Xfour\Xthree\Xthree\Xone$
&$2\times\ \Xseven\Xfour\Xfour\Xtwo\Xone$
&$2\times\ \Xseven\Xfive\Xthree\Xtwo\Xone$
&$\Xseven\Xfive\Xthree\Xone\Xone\Xone$\\
&&&&&\\
$\Xseven\Xthree\Xthree\Xtwo\Xtwo\Xone$
&$\Xseven\Xfour\Xtwo\Xtwo\Xtwo\Xone$
&$\Xseven\Xfive\Xfour\Xone\Xone$
&$\Xseven\Xsix\Xthree\Xone\Xone$
&$\Xseven\Xthree\Xthree\Xthree\Xone\Xone$
&$\Xseven\Xfour\Xthree\Xtwo\Xone\Xone$\\
&&&&&\\
$\Xn8\Xfour\Xthree\Xthree$
&$\Xn8\Xfour\Xfour\Xtwo$
&$\Xn8\Xfive\Xthree\Xtwo$
&$\Xn8\Xthree\Xthree\Xtwo\Xtwo$
&$2\times\ \Xn8\Xfour\Xtwo\Xtwo\Xtwo$
&$\Xn8\Xfive\Xfour\Xone$\\
&&&&&\\
$\Xn8\Xsix\Xthree\Xone$
&$\Xn8\Xthree\Xthree\Xthree\Xone$
&$2\times\ \Xn8\Xfour\Xthree\Xtwo\Xone$
&$\Xn8\Xfive\Xtwo\Xtwo\Xone$
&$\Xn8\Xthree\Xtwo\Xtwo\Xtwo\Xone$
&$\Xn8\Xfive\Xthree\Xone\Xone$\\
&&&&&\\
$\Xn8\Xthree\Xthree\Xtwo\Xone\Xone$
&$\Xn8\Xfour\Xtwo\Xtwo\Xone\Xone$
&$\Xn8\Xfour\Xthree\Xone\Xone\Xone$
&$\Xn9\Xfive\Xtwo\Xtwo$
&$\Xn9\Xthree\Xtwo\Xtwo\Xtwo$ 
&$\Xn9\Xfour\Xfour\Xone$ \\
&&&&&\\
$\Xn9\Xthree\Xthree\Xtwo\Xone$
&$\Xn9\Xfour\Xtwo\Xtwo\Xone$
&$\Xn9\Xfour\Xthree\Xone\Xone$
&$\Xn9\Xfive\Xtwo\Xone\Xone$
&$\Xn{11}\Xone\Xone\Xone\Xone\Xone\Xone\Xone$ & \\
&&&&&\\
\hline
\end{tabular}
\end{center}

\section{Some glueballs of $E_6$}
In general, we do not understand how glueballs are mapped.
For example, in the basic example of $SU(N_c)\leftrightarrow 
SU(N_f-N_c)$ of \cite{Seiberg:1995pq},
the spinorial glueball superfields do not appear to have 
partners in the dual theory~\footnote{This is actually not
a problem because these invariants, although chiral, 
are not primary: they 
are descendants~\cite{Berkooz:1996cb}. 
We will not try to discriminate between primaries and
descendants in this appendix. We thank Andreas Karch and
Micha Berkooz for this comment.}: 
\begin{eqnarray}
W_\alpha Q^{N_c} =\ \Xn{N_c-1}\Xone\ \nonumber\ ,
\end{eqnarray}
in the electric theory, clearly transforms differently from:
\begin{eqnarray}
\tilde W_\alpha q^{N_f-N_c} =\ \XBn{N_f-N_c-1}\XBone\ =\  
\Xn{N_c-1}\XN{N_f-1}\ \nonumber\  ,
\end{eqnarray}
in the magnetic theory.

For $E_6$, the situation is of course much worse.
We find that there is a rather large number of glueballs for $E_6$, 
the invariants involving the 78-dimensional adjoint representation 
for the glueball superfield $W_\alpha$.
First, the
Lorentz-spinor invariants of the form $W_{\alpha} Q^{3n}$:
\begin{eqnarray}
J_1 &= \78 \27^3_{[1^2\,1]}=\ \Ytwo\Yone  
& J_2 = \78 \27^6_{[1^3\,1^2\,1]}=\ \Ythree\Ytwo\Yone \nonumber\ \\
J_3 &= \78 \27^6_{[1^4\,1^2]}=\ \Yfour\Ytwo 
&J_4 = \78\27^9_{[1^4\,2^2\,1]}=\ \Yfour\Ytwo\Ytwo\Yone \nonumber\ \\
J_5 &= \78\27^9_{[1^4\,1^3\,1^2]}=\ \Yfour\Ythree\Ytwo 
&J_6 = \78\27^9_{[1^5\,2^2]}=\ \Xfive\Xtwo\Xtwo\nonumber\ \\
J_7 &= \78\27^9_{[1^5\,1^3\,1]}=\ \Xfive\Xthree\Xone
&J_8 = \78\27^9_{[1^5\,1^2\,2]}=\ \Xfive\Xtwo\Xone\Xone\nonumber\ \\
J_9 &= \78\27^9_{[1^6\,1^2\,1]}=\ \Xsix\Xtwo\Xone \nonumber\ \\
J_{10} &= \78\27^{12}_{[2^4\,1^3\,1]}=\ \Yfour\Yfour\Ythree\Yone
&J_{11} = \78\27^{12}_{[1^5\,1^4\,1^3]}=\ \Xfive\Xfour\Xthree \nonumber\ \\
J_{12} &= \78\27^{12}_{[1^5\,1^3\,2^2]}=\ \Xfive\Xthree\Xtwo\Xtwo 
&J_{13} = \78\27^{12}_{[1^5\,3^2\,1]} =\  \Xfive\Xtwo\Xtwo\Xtwo\Xone 
\nonumber\ \\
J_{14} &= \78\27^{12}_{[1^6\,3^2]}=\ \Xsix\Xtwo\Xtwo\Xtwo 
&J_{15,16} = \78\27^{12}_{[1^6\,1^3\,1^2\,1]}=\ \Xsix\Xthree\Xtwo\Xone \nonumber\ \\
J_{17} & = \78\27^{12}_{[1^6\,1^4\,2]}=\ \Xsix\Xfour\Xone\Xone 
&J_{18} = \78\27^{12}_{[1^6\,2^2\,2]}=\ \Xsix\Xtwo\Xtwo\Xone\Xone 
\nonumber\ \\
J_{19} &= \78\27^{12}_{[1^6\,1^3\,3]}=\ \Xsix\Xthree\Xone\Xone\Xone 
&J_{20} = \78\27^{12}_{[1^7\,1^4\,1]}=\ \Xseven\Xfour\Xone \nonumber\ \\
J_{21} &= \78\27^{12}_{[1^7\,1^3\,1^2]}=\ \Xseven\Xthree\Xtwo 
&J_{22,23,24} = \78\27^{12}_{[1^7\,1^2\,3]}=\ \Xseven\Xtwo\Xtwo\Xone 
\nonumber\ \\
J_{25,26} &= \78\27^{12}_{[1^7\,1^3\,2]}=\ \Xseven\Xthree\Xone\Xone
&J_{27} = \78\27^{12}_{[1^8\,1^3\,1]}=\ \Xn8\Xthree\Xone\nonumber\ \\
J_{28} &= \78\27^{12}_{[1^8\,1^2\,2]}=\ \Xn8\Xtwo\Xone\Xone 
&J_{29} = \78\27^{12}_{[1^8\,4]}=\ \Xn8\Xone\Xone\Xone\Xone \nonumber\ .
\end{eqnarray}
Then the Lorentz-scalar invariants of the form $W_{\alpha}^2 Q^{3n}$:
\begin{eqnarray}
K_1 &=&\78^2\27^3_{[3]}=\ \Yone\Yone\Yone \qquad 
K_2=\78^2\27^3_{[1^2\,1]}=\ \Ytwo\Yone \nonumber\  \\
K_3 &=& \78^2\27^3_{[1^2\,1]}=\ \Ytwo\Yone \qquad
K_4 = \78^2\27^3_{[1^3]}=\ \Ythree \nonumber\ \\
K_{5,6} &=&  \78^2\27^6_{[2^3]}=\ \Ythree\Ythree \qquad
K_{7,8} = \78^2\27^6_{[3^2]}=\ \Ytwo\Ytwo\Ytwo \nonumber\ \\
K_9 &=& \78^2\27^6_{[2^2\,2]}=\ \Ytwo\Ytwo\Yone\Yone \qquad
K_{10,11,12,13}= \78^2\27^6_{[1^4\,1^2]}=\ \Yfour\Ytwo \nonumber\ \\
K_{14,15,16,17,18} &= & \78^2\27^6_{[1^3\,1^2\,1]}=\ 
\Ythree\Ytwo\Yone \qquad
K_{19,20} = \78^2\27^6_{[1^4\,2]}=\ \Yfour\Yone\Yone\nonumber\ \\
K_{21} &=&  \78^2\27^6_{[1^5\,1]}=\ \Xfive\Xone \qquad 
K_{22,23}= \78^2\27^6_{[1^3\,3]}=\ \Ythree\Yone\Yone\Yone\nonumber\ \ . 
\end{eqnarray}
Due to this proliferation of glueball invariants, we stop the list here. 


\newpage

\section{Computer Program Code for LiE}
The group theory program LiE  can be obtained for free from a 
variety of sources, e.g. \texttt{
http://wwwmathlabo.univ-poitiers.fr/$\tilde{\ } $maavl/LiE/\ .
}
It is written in $C$, easy to install and easy to use.
It comes with a well-written and useful manual~\cite{LiE:1996}.
(There is another well-known group theory software, Schur. However it
is not free.)
We include some commented code below.
\vskip.5in

\noindent
0 \#Write to a default file named `monfile' \\
\texttt{1 on monitor \\
2 maxobjects 2000000 \\
3 maxnodes 12000 \\
4 thegroup = E6; therank = 6;}\\
5 \#n is the degree of the invariants we are computing\\
\texttt{
6 n=9;\\
7 setdefault Lie\_group(1,n);\\
8 fund = [1,0,0,0,0,0,0,0,0];\\
9 y= poly\_null(n)\\
10 ii=0;\\
}
11 \#The next line iterates over all the partitions of n=6\\
\texttt{
12 for lambda row partitions(n) do}\\
13 \#Compute the plethysm of degree $n$ of $E_6$\\
\texttt{
14 b=plethysm(lambda,[1,0,0,0,0,0],thegroup);}\\
15 \#Compute the $SU(n)$ representation for the partition $\lambda$\\
\texttt{
16 x=plethysm(lambda,fund);}\\
  \# Line 16 is very slow. More efficient is to use the built-in 
function \texttt{from\_part} \\
  \# (after adding a zero to each partition to get 
representations of $A_n$ instead of $A_{n-1}$). 
17 \#Print useful information at each iteration\\
\texttt{
18 ii=ii+1;\\
19 print("partition number:"); print(ii);\\
20 print("partition:"); print(x); print(lambda);\\
21 print("number of objects used:"); print(used);\\
22 print("the plethysm:");\\
23 print(b);\\
24 print("Multiplying by W\_alpha:");\\
}
25 \#Tensor the plethysm $b$ with the adjoint\\
26 \#The coefficient \texttt{1X} is for compatibility with 
$b$ because $b$ is a polynomial\\
\texttt{
27 cc=tensor(1X[0,1,0,0,0,0],b,thegroup);\\
28 print(cc);\\
}
29 \#Pick the singlet(s) from the final expression\\
\texttt{ 
30 a=cc[1];}\\
31 \#Check that $a$ is a singlet; if so, add it to $y$\\
\texttt{
32 if (a==(coef(a,1)* poly\_one(therank))) then \\
33 y= y+( coef(a,1)*x);\\
34 print("This is the coefficient:");\\
35 print(coef(a,1)*x);\\
36 print("This is 'y': "); print(y);}\\
37 \# 'y' is the full list of invariants \\
38 \#Closes the \texttt{if}\\
\texttt{
39 fi; \\
}
40 \#Activate the garbage collection:\\
41 \#Essential memory management!\\
\texttt{
42 gcol; \\
}
43 \#Close the \texttt{do} loop\\
\texttt{
44 od;\\
}
45 \#Define the previously found lower degree invariants\\
\texttt{
46 j1 = plethysm([2,1],fund);\\
47 j2 = plethysm([3,2,1],fund);\\
48 j3 = plethysm([2,2,1,1],fund);\\
49 \\
50 r1 = plethysm([3],fund);\\
51 r2 = plethysm([2,2,2],fund);\\
52 r3 = plethysm([3,3,1,1,1,1],fund);\\
53 r4 = plethysm([3,3,3,3],fund);\\
54 r5 = plethysm([4,4,1,1,1,1],fund);\\
55 r6 = plethysm([4,2,2,1,1,1,1],fund);\\
}
56 \#Compute the product of lower degree invariants among themselves\\
\texttt{
57 f = tensor(r1,j2+j3)+tensor(sym\_tensor(2,r1),j1)+tensor(r2,j1);\\
58 print("This is f: ");\\
59 print(f);\\
60 print("These are the new fundamental invariants: ");\\
61 print(y-f);\\
}

\end{document}